\documentclass[
preprint,
preprintnumbers,
nofootinbib,
 amsmath,amssymb,
 aps, prl, showkeys,
]{revtex4-2}

\usepackage{float}
\usepackage[normalem]{ulem}
\DeclareMathAlphabet{\mathpzc}{OT1}{pzc}{m}{it}
\usepackage{xcolor}
\usepackage[normalem]{ulem} 

\usepackage{mathtools}
\usepackage{color}

\usepackage{tabularx,ragged2e,booktabs,caption}

\usepackage[colorlinks,citecolor=black,urlcolor=blue,bookmarks=false,hypertexnames=true]{hyperref} 

\usepackage{bm}
\usepackage{amsmath}
\usepackage{amsfonts, verbatim}
\usepackage{amsthm}             
\usepackage{amssymb}            
\usepackage{bbm}                
\usepackage{booktabs}           
\usepackage{graphicx}
\usepackage{subcaption}
\usepackage{wrapfig}
\usepackage{float}
\usepackage[normalem]{ulem}
\DeclareMathAlphabet{\mathpzc}{OT1}{pzc}{m}{it}
\usepackage{xcolor}
\usepackage[normalem]{ulem} 
\usepackage{mathtools}
\usepackage{color}
\usepackage{graphicx}
\usepackage{setspace}

\usepackage{tabularx,ragged2e,booktabs}
\usepackage[font=small,labelfont=bf]{caption}
\usepackage{enumerate}
\usepackage{multirow}
\usepackage{mathtools}

\usepackage{algorithm}
\usepackage{algpseudocode}
\newtheorem{theorem}{Theorem}[section]

\newtheorem{remark}[theorem]{Remark}

\newcommand{\mbf}[1]{\boldsymbol{#1}}

\newcommand{\dotp}[1]{\langle{#1}\rangle}

\newcommand{\abs}[1]{\big| #1 \big|}

\newcommand{\norm}[1]{\left\| #1 \right\|}
\newcommand{\R}{\mathbb{R}}

\newcommand{\br}{\mbf{r}}

\newcommand{\bv}{\mbf{v}}
\newcommand{\bV}{\mbf{V}}

\newcommand{\bx}{\mbf{x}}
\newcommand{\bX}{\mbf{X}}

\newcommand{\mE}{\mathcal{E}}

\newcommand{\mH}{\mathcal{H}}

\newcommand{\mL}{\mathcal{L}}
\newcommand{\mO}{\mathcal{O}}

\newcommand{\mS}{\mathcal{S}}

\newcommand{\bigO}{\mathcal{O}}
\newcommand{\idxcl}{k}

\newcommand{\intkernel}{\phi}
\newcommand{\bintkernel}{{\bm{\phi}}}

\newcommand{\lintkernel}{\hat\intkernel}
\newcommand{\blintkernel}{{\widehat{\bm{\intkernel}}}}
\newcommand{\intkernelvar}{\varphi}
\newcommand{\bintkernelvar}{{\bm{\varphi}}}
\newcommand{\basis}{\psi}

\newcommand{\rhsf}{\mathbf{f}}
\newcommand{\hypspace}{\mH}

\DeclareMathOperator*{\argmin}{argmin}

\begin{document}


\title{Machine Learning for Discovering Effective Interaction Kernels between Celestial Bodies from Ephemerides}


\author{Ming Zhong}
 \email{mingzhong@tamu.edu}
 \affiliation{Department of Applied Mathematics \& Statistics,
 Johns Hopkins University}
\author{Jason Miller}
 \affiliation{Department of Applied Mathematics \& Statistics,
 Johns Hopkins University}
\author{Mauro Maggioni}
  \affiliation{Department of Mathematics, 
  Department of Applied Mathematics \& Statistics,
  Johns Hopkins University}

\date{\today}

\begin{abstract}
Building accurate and predictive models of the underlying mechanisms of celestial motion has inspired fundamental developments in theoretical physics. Candidate theories seek to explain observations and predict future positions of planets, stars, and other astronomical bodies as faithfully as possible. We use a data-driven learning approach, extending that developed in Lu et al. ($2019$) and extended in Zhong et al. ($2020$), to a derive stable and accurate model for the motion of celestial bodies in our Solar System. Our model is based on a collective dynamics framework, and is learned from the NASA Jet Propulsion Lab's development ephemerides.  By modeling the major astronomical bodies in the Solar System as pairwise interacting agents, our learned model generate extremely accurate dynamics that preserve not only intrinsic geometric properties of the orbits, but also highly sensitive features of the dynamics, such as perihelion precession rates. Our learned model can provide a unified explanation to the observation data, especially in terms of reproducing the perihelion precession of Mars, Mercury, and the Moon.  Moreover, Our model outperforms Newton's Law of Universal Gravitation in all cases and performs similarly to, and exceeds on the Moon, the Einstein-Infeld-Hoffman equations derived from Einstein's theory of general relativity.
\end{abstract}

\maketitle


%
\section{Introduction}
Precise modeling and prediction of celestial motion has provided the motivation for many early developments in mathematics and physics \cite{Longair, Roy}.  These discoveries of fundamental physics usually worked hand in hand with the development of novel mathematical tools to provide explanations of data from observations.   

A major breakthrough in the rigorous theory for celestial motion was published by Copernicus in $1543$. In $1609$, Kepler improved this theory by publishing his laws on planetary motion.  Ultimately, it took the development of calculus to build a complete model of celestial dynamics.  In $1687$,  Newton presented the famous $\frac{1}{r^2}$-form of the law of gravity. However, in $1845$, Le Verrier discovered that the perihelion precession rate of Mercury could not be fully explained by Newton's theory of gravity.   It took another $70$ years, and the development of Riemannian geometry and relativity, for Einstein to explain that the discrepancy was due to the effect of the curvature of spacetime around the Sun.  Einstein's theory has since been applied to celestial motion well beyond the Solar System (see \cite{Turyshev, Bambi2018, PATTON, PhysRevDLong} and references therein).  

With the rapid development of advanced observation technologies,  statistics and machine learning have enabled us to analyze big data sets and discover novel patterns that are nearly impossible for a human to identify.  It is well suited to play a complementary role to traditional physical reasoning in the pursuit of discovering fundamental physics \cite{Jordan255, RevModPhys.91.045002}.   There has been extensive research applying machine learning in science (especially in physics) and engineering, examples include: learning PDEs \cite{Bar-Sinai,Schaeffer6634}, governing equations \cite{Champion}, behavior in biology  \cite{Chiel993}, and fluid mechanics \cite{Raissi1026, Han219}. Further examples include: many-body problems in quantum systems \cite{Carleo602}, mean field games \cite{Ruthotto}, meteorology \cite{Ham2019}, dynamical systems \cite{Costa1501, Brunton3932, YairE, Bongard9943}, and discrete field theory \cite{Qin2020}.

By combining effective theory \cite{Wells} and machine learning, our data-driven modeling can provide interpretable and meaningful physical models which can be used to model observations with extremely high accuracy, preserving not only the geometric properties of the trajectories, but also localized dynamical features such as perihelion precession rates.  Our approach is developed based on a collective dynamics framework, derived from classical Lagrangian mechanics.  We begin with a second order system in the form
\begin{equation}\label{eq:second_order}
m_i\ddot\bx_i(t) = \sum_{i' = 1}^N \frac{1}{N} \intkernel(\norm{\bx_{i'}(t) - \bx_i(t)})(\bx_{i'}(t) - \bx_i(t)),
\end{equation}
for $i = 1, \ldots, N$.  Here, $\intkernel: \R^+ \rightarrow \R$, is known as an \textbf{interaction kernel}.  The problem of inferring the interaction kernel from observed trajectories, in a non-parametric fashion, was considered in \cite{BFHM17}, and extended in multiple directions (both of theoretical and practical relevance) in \cite{LZTM2019, ZMM2020, MTZM2020, MMQZ2021} (see the section ``Related Works'' in SI for a detailed discussion).

In this work, we consider the problem of inferring the interaction kernels of celestial bodies in the Solar System from trajectory data, with minimal a priori knowledge about their form; in particular we assume no knowledge of geometric properties of the trajectories (e.g. elliptical, closed, etc.), of masses of the celestial bodies (in fact, not even the concept of mass), and no assumption on the form of the interaction kernels (e.g. inverse powers of pairwise distance).  We are particularly interested in discovering effective models of gravitation that best explain empirical data (both from observations and simulations) and compare these effective models to celebrated models from physics.  We use trajectory data from the Jet Propulsion Laboratory's (JPL) development ephemerides.  We compare the performance of our model to the JPL data, as well as to two important models: one based on Newton's Universal Law of Gravitation, and the other is a first-order approximation of Einstein's general relativity theory, namely the Einstein-Infeld-Hoffman (EIH) model.  We discover that our model can provide superior performance over the other two models in terms of trajectory error in comparing the position and velocity data,  preserving the geometric properties (period/aphelion/perihelion) of the trajectories, and reproducing the highly sensitive and localized perihelion precession rates of three prototypical bodies: Mars (observation of its orbits led to classical Newtonian gravity), Mercury (where the general relativity effect is prominent), and the Moon (where gravity alone cannot provide a full explanation of the precession), see table \ref{tab:PR} for details.
\section{Results}
\begin{figure}[H]
\centering
\includegraphics[width=0.7\textwidth]{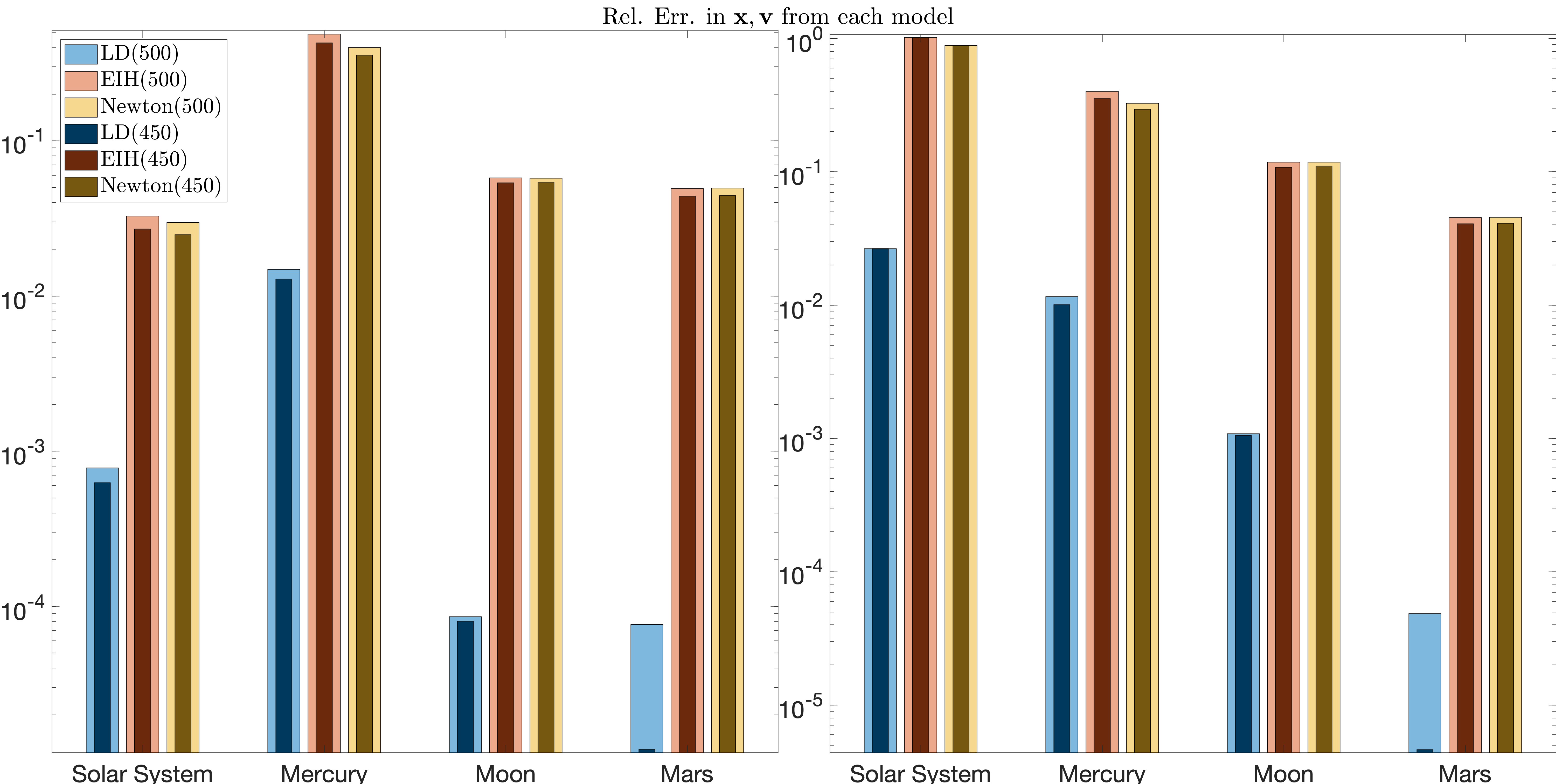} 
\caption{Relative errors in position ($\bx$) and velocity ($\bv$) from each model (LD/EIH/Newton) compared to the JPL observation data (using \eqref{eq:sys_traj_err} and \eqref{eq:AO_traj_err}) for the full Solar System, Mars, Mercury, and the Moon over $450$ and $500$ year trajectories.  The errors over $450$ years have narrower width and darker color, and are laid on top of the errors over $500$ years, which have broader width and a lighter color.  Different colors correspond to different models: dark/light blue for LD, dark/light red for EIH, and dark/light yellow for Newton. Our learned model demonstrates high accuracy in terms of trajectory error in all cases.}
\label{fig:JPL_trajErr}
\end{figure} 
We present the most significant results from our machine-learning procedure for celestial dynamics (labeled as LD for ``Learned Dynamics'') compared to the JPL observation data (JPL), the EIH model (EIH), and the Newton model (Newton). The errors in comparing position and velocity are shown in Figure \ref{fig:JPL_trajErr}.  LD demonstrates superior performance in terms of recreating the trajectories, on both training (a period of $450$ years) and testing (a subsequent period of $50$ years) data.  Table \ref{tab:PR} shows the perihelion precession rates (PPRs) of Mars, Mercury, and the Moon from the four different models considered, estimated over $450$ years of trajectory data.
\begin{figure}[H]
\centering
\includegraphics[width=0.7\textwidth]{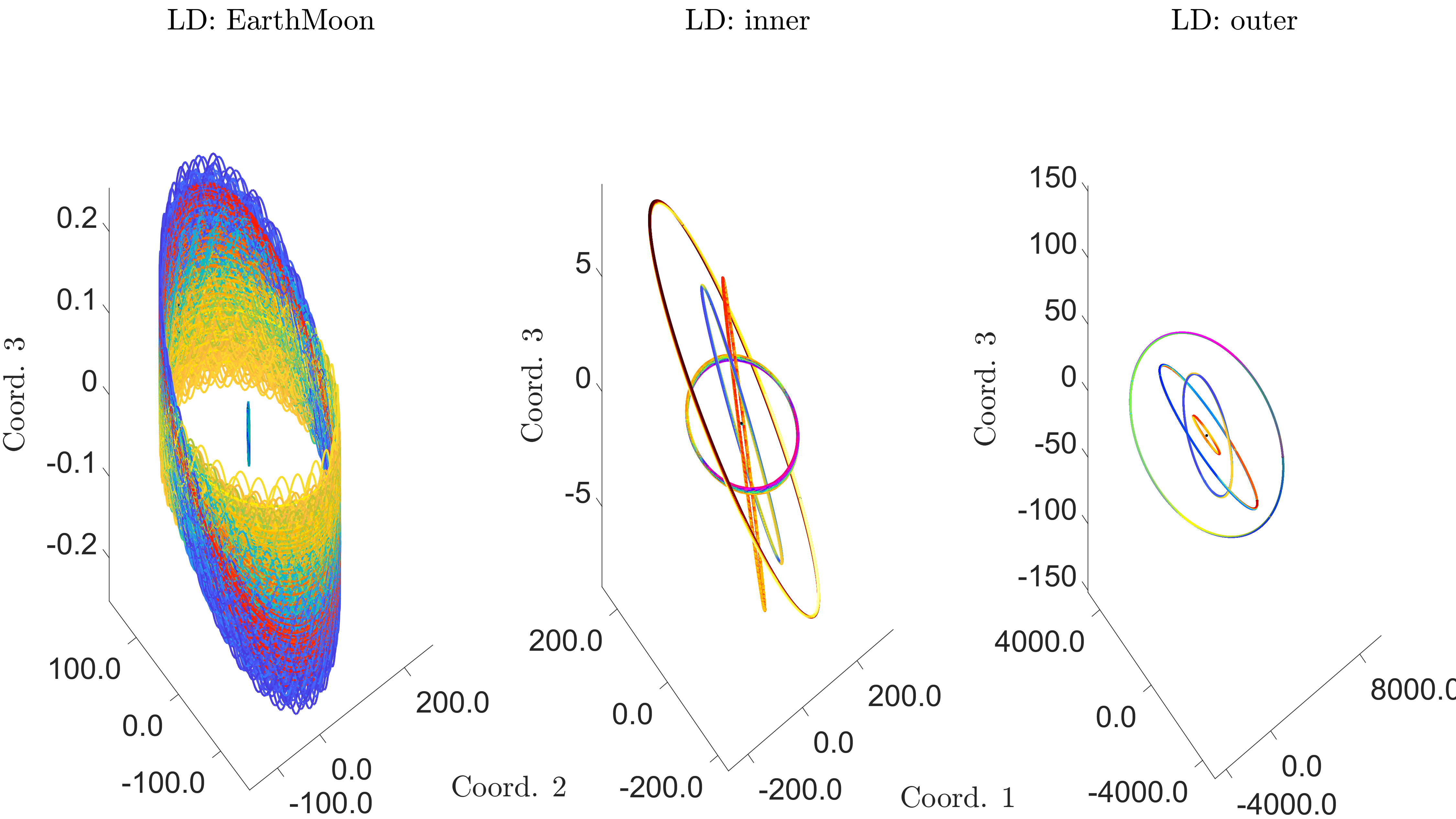}
\caption{Trajectories of the Solar System from the learned model (LD) evolved over $500$ years with the initial position/velocity taken at year $1500$ from the JPL data. \text{Left}: Earth-Moon-Sun system; \textbf{Middle}: Inner Solar System; \textbf{Right}: Outer Solar System.  Since we are maintaining $10^{-3}$ relative accuracy at reproducing the positions for the celestial bodies in our Solar System, we do not show the visually indistinguishable trajectories from the JPL data.}
\label{fig:JPL_traj}
\end{figure} 
\begin{table}[H]
\centering
 \begin{tabular}{c | c | c | c | c } 
 \hline
         & JPL                 & LD                  & EIH                 & Newton             \\ 
 \hline
 Mars    & $1.52 \cdot 10^{3}$ & $1.50 \cdot 10^{3}$ & $1.52 \cdot 10^{3}$ & $1.52 \cdot 10^{3}$ \\         
 \hline
 Mercury & $576.58$            & $567.28$            & $569.48$            & $533.35$            \\
 \hline
 Moon    & $3.43 \cdot 10^{7}$ & $3.49 \cdot 10^{7}$ & $3.07 \cdot 10^{7}$ & $2.80 \cdot 10^{7}$ \\
 \hline
 \end{tabular}
 \caption{Perihelion precession rate (PPR) estimation for $3$ different celestial bodies from $4$ different models.  The algorithm for calculating the PPR is presented in Algorithm $1$ in Section ``Estimating Planet Information'' in SI. Our learned model reproduces in a consistent and accurate fashion the precession rate of all $3$ celestial bodies (we consider the JPL model as providing the ground truth); in particular it is substantially more accurate than the rate reproduced by the Newton model on Mercury, where it is nearly as accurate as the rate from the EIH model, and it is the most accurate at reproducing the complicated precession dynamics of the Moon.}
 \label{tab:PR}
\end{table}
Our learning procedure is able to capture the essence of the PPR, which is highly sensitive, localized, and representative of the characteristics of the dynamics for $3$ prototypical celestial bodies, with the fewest assumptions made on the observation data.  In Figure \ref{fig:JPL_traj}, we show the dynamics evolved using our learned models for $500$ years using a symplectic integrator described in the section ``Symplectic Integration'' in SI.
\section{Model Description}\label{sec:models}
We make the following assumptions in order to simplify the discussion: $i)$ we work with absolute time and space, which is valid in the low energy/low velocity setting of the Solar System; $ii)$ the gravitational mass is the the same as the inertial mass in the equations of motion; $iii)$ the gravitational effect acts instantaneously at any distance.  Improvements incorporating relativistic principles into our models is a focus of future works.
In our machine learning based approach we assume that the gravitational force depends only on pairwise distance, which is consistent with translation and rotation-invariance properties of the force; we do not assume any analytic expression for the gravitational force, nor do we assume knowledge of the concept mass, nor of how it affects the interactions. We work with the assumption that each celestial body is of its own ``type'', i.e. with its own type of gravitational force with other celestial bodies
 
Using the framework of Lagrangian Mechanics, we consider the Lagrangian, $\mL(t)$, of a closed Solar System of $N$ celestial bodies, each identified with its center of mass: 
\[
\mL(t) = \sum_{i = 1}^N\frac{1}{2}m_i\norm{\bv_i(t)}^2 - \frac{1}{2}\sum_{i, i' = 1}^N U_{i, i'}(\norm{\bx_{i'}(t) - \bx_i(t)}).
\]
Here $\bx_i,\bv_i \in \R^d$ ($d = 3$) are the position, velocity of the $i^{th}$ celestial body, $\norm{\cdot}$ is the usual Euclidean norm, and $U_{i, i'}: \R^+ \rightarrow \R$ is a potential energy depending only on pairwise distance, and it is parameterized by the unknown masses of celestial body $i$ and $i'$ in an unknown, possibly non-linear way.  We further assume that $U_{i, i'} = U_{i', i}$ and $U_{i, i} \equiv 0$.  Then, via the Lagrange equation, $\frac{d}{dt}\Big(\frac{\partial\mL}{\partial\bv_i}\Big) = \frac{\partial\mL}{\partial \bx_i}$, we arrive at the equation of motion for the $i^{th}$ celestial body:
\[
m_i\dot\bv_i(t) = \sum_{i' = 1}^N (U'_{i, i'}(\norm{\bx_{i'}(t) - \bx_i(t)}))\cdot\frac{\bx_{i'}(t) - \bx_i(t)}{\norm{\bx_{i'}(t) - \bx_i(t)}}.
\]
\begin{remark}
In the case of Newton's gravitational potential, we have $U_{i, i'}(\norm{\bx_{i'}(t) - \bx_i(t)}) = \frac{-Gm_im_{i'}}{\norm{\bx_{i'}(t) - \bx_i(t)}}$.  Newton came to the conclusion of this particular form based on Kepler's laws and the assumption that the gravitational force should have a ``$\frac{1}{r^p}$'' form with $p = 2$ being the only solution for closed elliptical orbits.
\end{remark}
Simplifying, we obtain the following equations of motion for the $i^{th}$ celestial body ($i = 1, \ldots, N$),
\begin{equation}\label{eq:first_model}
\dot\bv_i(t) = \sum_{i' = 1, i' \neq i}^N \intkernel_{i, i'}(r_{i, i'}(t))\br_{i, i'}(t).
\end{equation}
Here $\br_{i, i'}(t) \coloneqq \bx_{i'}(t) - \bx_i(t)$,  $r_{i, i'}(t) \coloneqq \norm{\br_{i, i'}(t)}$, and $\bintkernel:=[\intkernel_{i, i'}]_{i, i' = 1}^N$ is a set of \textbf{interaction kernels}, where each $\intkernel_{i, i'}$ also subtends information about the masses of celestial body $i$ and $i'$. From the above, the true form of $\intkernel_{i, i'}$ is $\frac{U'_{i, i'}(r)}{m_ir}$.
\section{Learning Framework}\label{sec:framework}
Our machine learning approach is aware of the general form of the equations of motion \eqref{eq:first_model}, but otherwise makes no a priori assumptions on the geometry of trajectories (such as it being elliptical), has no knowledge of the role of mass in the equations of motion (nor, therefore, of the masses of the celestial bodies), nor makes assumptions, besides regularity, on the functional forms of the interaction kernels. The learning procedure, based on those in \cite{BFHM17, LZTM2019, ZMM2020, MTZM2020, MMQZ2021}, takes as input only a set of discrete-time trajectory data $\{\bx_i(t_l), \bv_i(t_l), \dot\bv_i(t_l)\}_{i, l = 1}^{N, L}$, for $T_0 = t_1 < \cdots < t_L = T$, for the celestial bodies in our Solar System. 
To simplify the presentation, we introduce vectorized notation as follows:  let $\bX_{t_l}$ (resp. $\bV_{t_l}$) be the column vector obtained by concatenating the column vectors $(\bx_i(t_l))_{i=1}^N$ (resp.: $(\bv_i(t_l))_{i=1}^N$),  and
\[
\rhsf_{\bintkernelvar}(\bX_{t_l}) \coloneqq \begin{bmatrix} \vdots \\ \sum_{i' = 1}^N \intkernelvar_{i, i'}(r_{i, i'}(t_l))\br_{i, i'}(t_l) \\ \vdots \end{bmatrix}.
\] 
Here $\bintkernelvar \coloneqq [\intkernelvar_{i, i'}]_{i, i' = 1}^N$.   Note that $\bX_{t_l},  \bV_{t_l}, \rhsf_{\bintkernelvar}(\bX_{t_l}) \in \mathbb{R}^{D}$, with $D \coloneqq Nd$.  We define the $\norm{\cdot}_{\mS}$ on $\R^D$ as $\norm{\bX}_{\mS}^2 \coloneqq \sum_{i = 1}^N\norm{\bx_i}^2$.
\subsection{Non-parametric Learning of Interaction Kernels}
To simplify the discussion, we take equispaced time points, i.e. $t_{l} - t_{l - 1} = t_{l + 1} - t_l$ for $l = 2, \ldots, L - 1$; however equispacing is not mandatory for our algorithm.  We find a set of estimated interaction kernels $\blintkernel \coloneqq [\lintkernel_{i, i'}]_{i, i' = 1}^N$ by minimizing the error functional
\begin{equation}\label{eq:loss_fun_1D}
\mE_L(\bintkernelvar) \coloneqq \frac{1}{L}\sum_{l = 1}^L\norm{\ddot\bX_{t_l} - \rhsf_{\bintkernelvar}(\bX_{t_l})}_{\mS}^2.
\end{equation}
Here,  $\bintkernelvar = [\intkernelvar_{i, i'}]_{i, i' = 1}^N$ with each $\intkernelvar_{i, i'} \in \hypspace_{i, i'}$ a compact (in the $L^\infty$-norm) and convex subset of square-integrable functions $L^2([R_{i, i'}^{\min}, R_{i, i'}^{\max}])$, where $R_{i, i'}^{\min/\max}$ is the minimum/maximum interaction radius for the interaction kernel, $\intkernel_{i, i'}$ when $i \neq i'$; when $i = i'$, we simply take $[R_{i, i'}^{\min}, R_{i, i'}^{\max}]$ to be $[0, 1]$.  Let $\bm\hypspace \coloneqq \oplus_{i, i' =1 }^N \hypspace_{i, i'}$ and $\blintkernel \coloneqq \argmin_{\bintkernelvar \in \hypspace} \mE_L(\bintkernelvar)$. The convergence of $\blintkernel$ to the true interaction kernels, as the number of observations increases, in the more restrictive setting of trajectories generated by random initial conditions, and known masses, is studied in \cite{MTZM2020}: one of the major takeaways of that analysis is that even if the learning problem is for a system in $D$ dimensions, upon choosing suitable hypothesis spaces of dimension growing with the number of observations (in line with classical nonparametric statistics), the learning rate only depends on the number of variables in the interaction kernel, which in this case is $1$ (pairwise distance), and is near-optimal.
\subsection{Performance Measures}
We consider two other types of performance measures, both of which depend on the difference in the observed planetary motion and the recreated planetary motion -- which is generated by evolving the dynamical system using \eqref{eq:first_model} our learned model of the interaction kernels, or the other two models (Newton or EIH) starting at the same initial conditions (IC) as the observation data.  Let $\bX_t$ be the observed positions of the $N$-body system for $t \in [T_0, T]$, and $\hat\bX_t$ be the recreated positions evolved from the same IC as the observed system with the learned interaction kernels or known equations of motion defined by the Newton or the EIH models over $t \in [T_0, T]$, then consider
\begin{equation}\label{eq:sys_traj_err}
\text{Err}_1 \coloneqq \frac{\max_{t \in [T_0, T]}\norm{\bX_t - \hat\bX_t}_{\mS}}{\max_{t \in [T_0, T]}\norm{\bX_t}_{\mS}}.
\end{equation}
However, by considering the system as a whole, errors in an individual celestial body's trajectory with positions with smaller norm could be obscured.  To avoid this, we consider and report a second type of error,
\begin{equation}\label{eq:AO_traj_err}
\text{Err}_{2, i} \coloneqq \frac{\max_{t \in [T_0, T]}\norm{\bx_i(t) - \hat\bx_i(t)}}{\max_{t \in [T_0, T]}\norm{\bx_i(t)}}
\end{equation}
We use similar formulas for computing estimation errors in velocities $\bV_t$ and $\bv_i(t)$.
\subsection{Computational and Numerical Aspects}
Let $n = n_{i, i'}$\footnote{We use a uniform $n$ for all $\hypspace_{i, i'}$ when $i \neq i'$, hence we suppress the dependence of $n$ on $(i, i')$.} denote the number of basis functions for the hypothesis space $\hypspace_{i, i'}$ when $i \neq i'$; when $i = i'$, we simply take $n_{i, i} = 1$. The total computational cost for solving the learning problem is $\bigO(LN^2 + LdN^2n^2 + N^3n^3)$.  The computational bottleneck comes in the variable $L$ when $L \gg Nn$, since we are processing hundreds of years of data.  However, we can parallelize our learning method in $L$ by splitting the long trajectory into pieces, which significantly reduces the computing time and storage.  The total data storage needed for the observation of $500$ years (i.e. $L \approx 500 \times 365$) of position/velocity data of $N = 10$ celestial bodies (i.e. $2LNd$) amounts to roughly $11\times10^6$ numbers, hence parallelization in $L$ is needed in order to compute the pairwise quantities efficiently.  Meanwhile,  during the assembly of the learning matrices, a matrix of size $\bigO(LdNn)$ is generated for each celestial body, which requires handling a total of roughly $4.4\times10^9$ numbers.  Hence, the assembly of the learning matrices also has to be done in parallel, see the section ``Computational Complexity'' in SI for detailed discussion.  Once the final matrix is assembled, it is of size $\bigO(N^2n^2)$ (with $n \approx 10^2$), and its inversion can be easily handled.

For integration on our learned model and the Newton model, we use a symplectic integrator (fourth order Leapfrog with $h = 10^{-2}$); and for the integration on the EIH model, we use MATLAB's fully implicit integrator, $\text{ode}15\text{i}$, with relative tolerance set at $10^{-8}$ and absolute tolerance at $10^{-11}$. 
\subsection{Modern Ephemerides}
We choose the National Aeronautics and Space Administration's (NASA) Jet Propulsion Laboratory's (JPL) Development Ephemeris (DE), numbered as DE$430/431$, as our only source of observation data.  These modern ephemerides are routinely updated and maintained, and it has been used in NASA's space exploration missions, and published by the Astronomical Almanac, since $1984$.  Details on DE$430/431$ can be found in \cite{SW2007}.
\section{Detailed Results and Performance}\label{sec:results}
\begin{figure}[H]
\centering
\begin{subfigure}[b]{0.48\textwidth} 
\centering
\includegraphics[width=\textwidth]{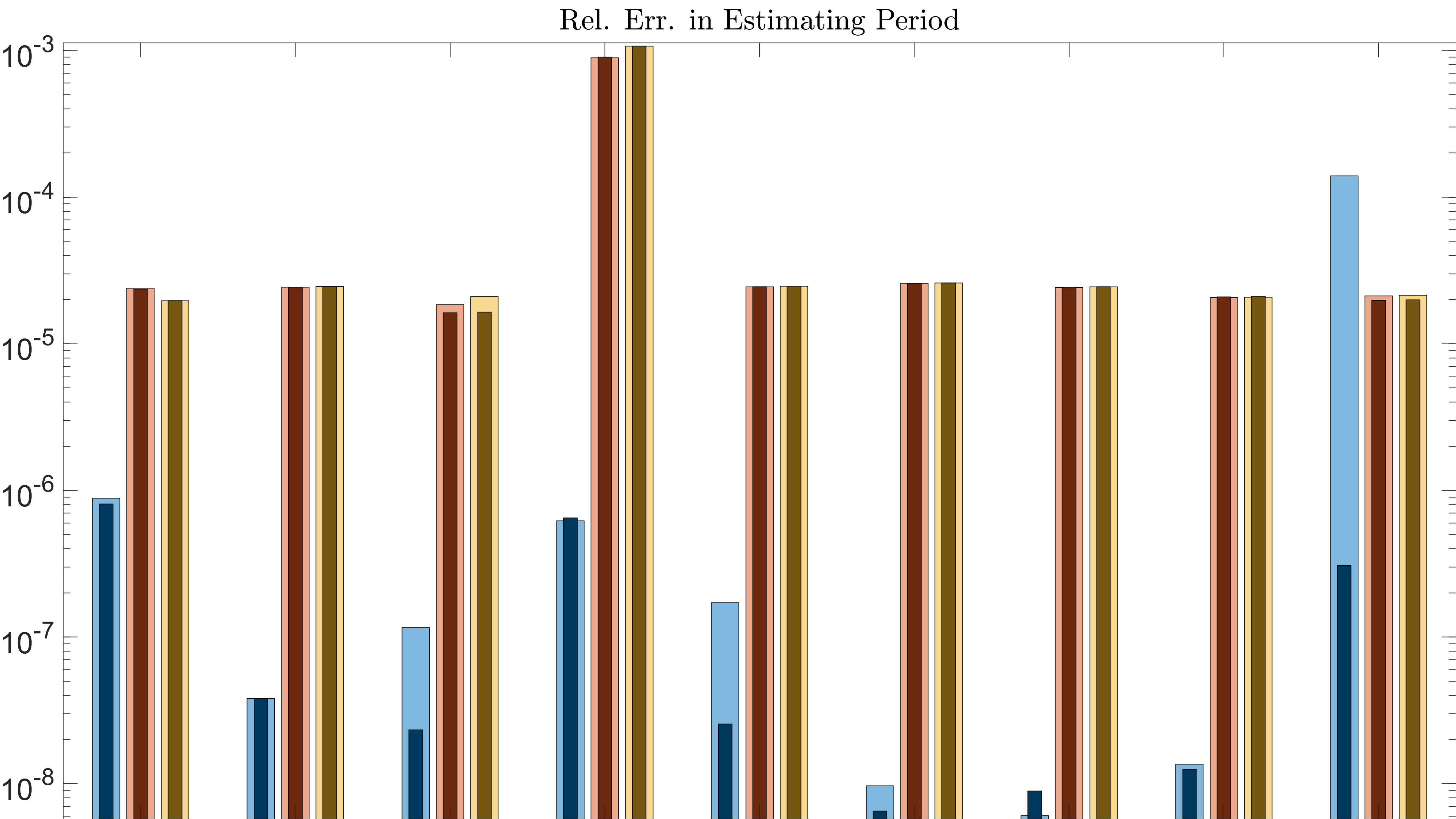} 
\end{subfigure} \\
\begin{subfigure}[b]{0.48\textwidth}
\centering
\includegraphics[width=\textwidth]{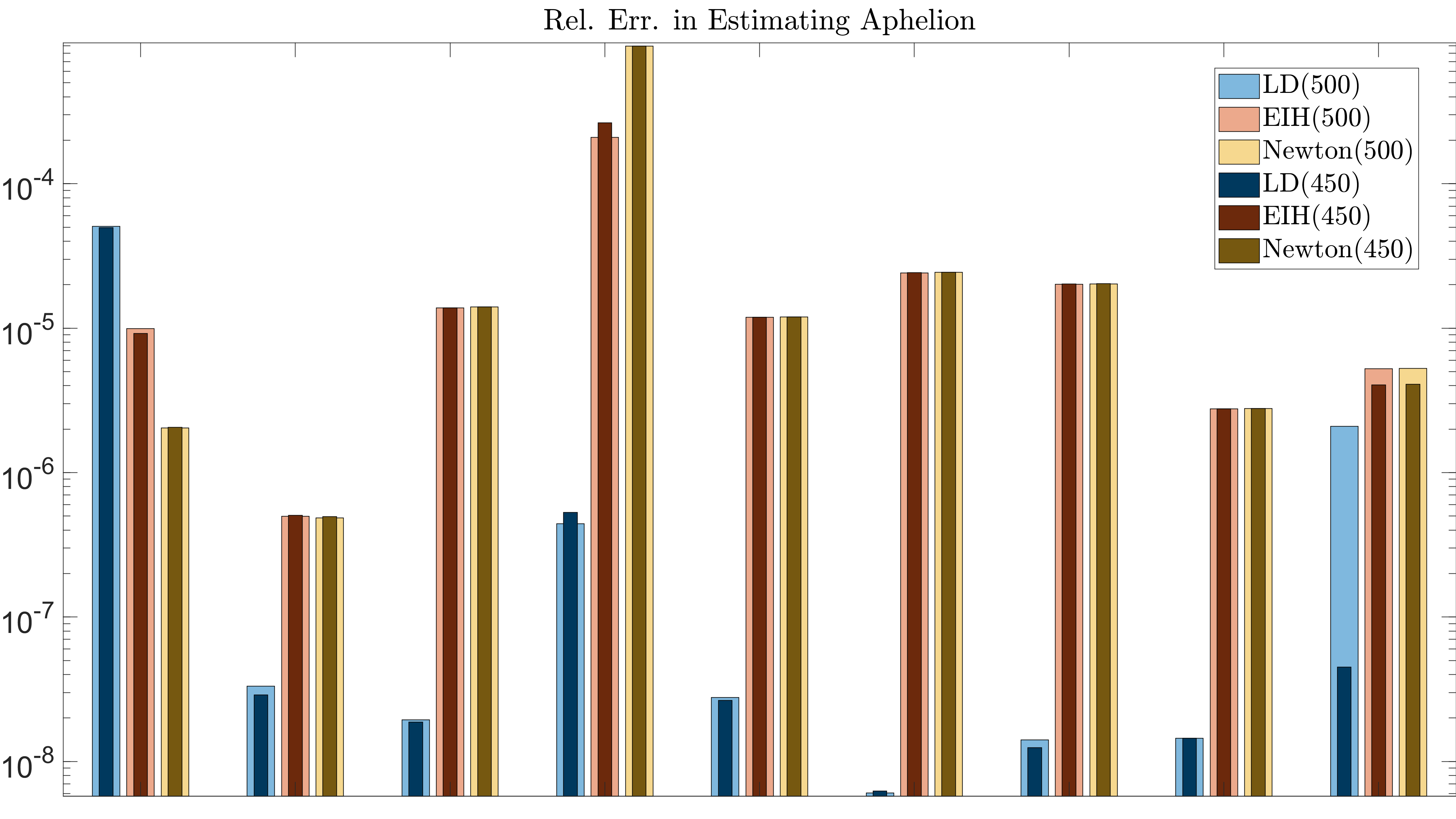}
\end{subfigure} \\
\begin{subfigure}[b]{0.48\textwidth}
\centering
\includegraphics[width=\textwidth]{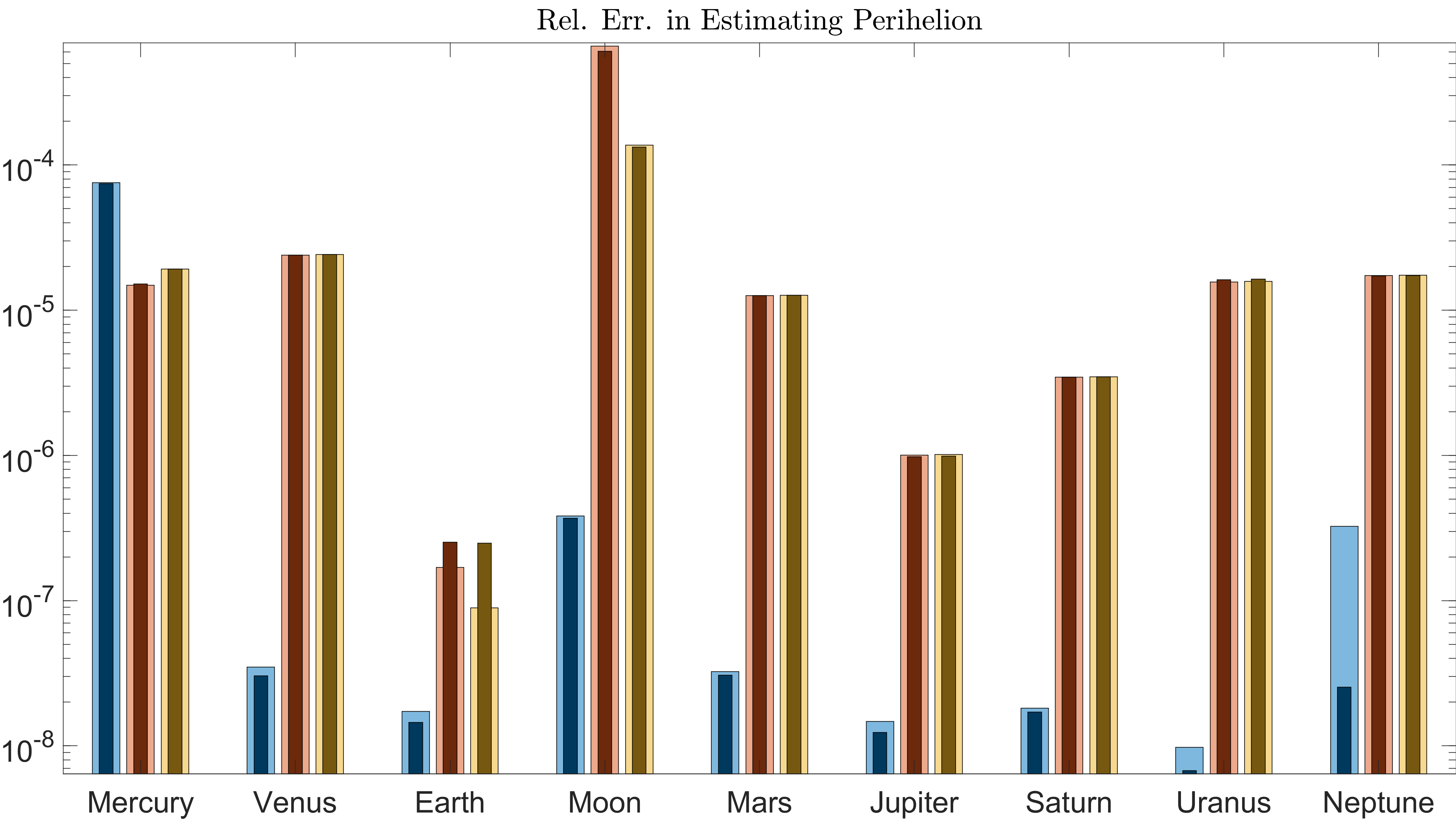}
\end{subfigure}
\caption{Comparison of relative errors in period/aphelion/perihelion from three different models (LD/EIH/Newton) compared to the JPL observation data for $9$ different celestial bodies over $450$ and $500$ year trajectories.  The errors over $450$ years have smaller width and darker color, and are laid on top of the errors over $500$ years, which have bigger width and lighter color.  Different colors correspond to different models: dark/light blue for LD, dark/light red for EIH, and dark/light yellow for Newton. The learned model demonstrates high accuracy in terms of trajectory error in almost all cases.}
\label{fig:JPL_PIerr}
\end{figure}

\begin{figure}[H]
\centering
\begin{subfigure}[b]{0.48\textwidth} 
\centering
\includegraphics[width=\textwidth]{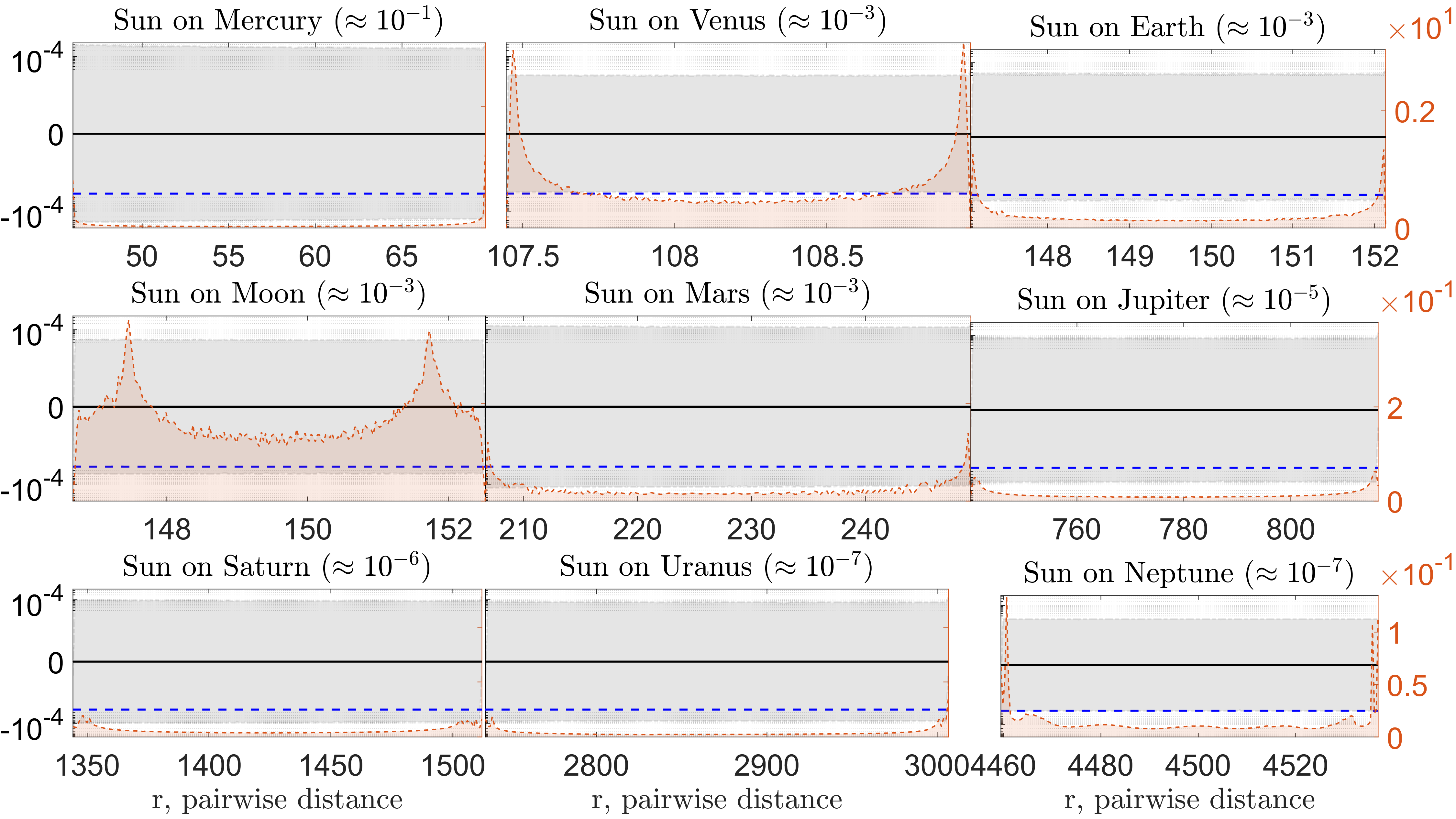} 
\end{subfigure}
\begin{subfigure}[b]{0.48\textwidth}
\centering
\includegraphics[width=\textwidth]{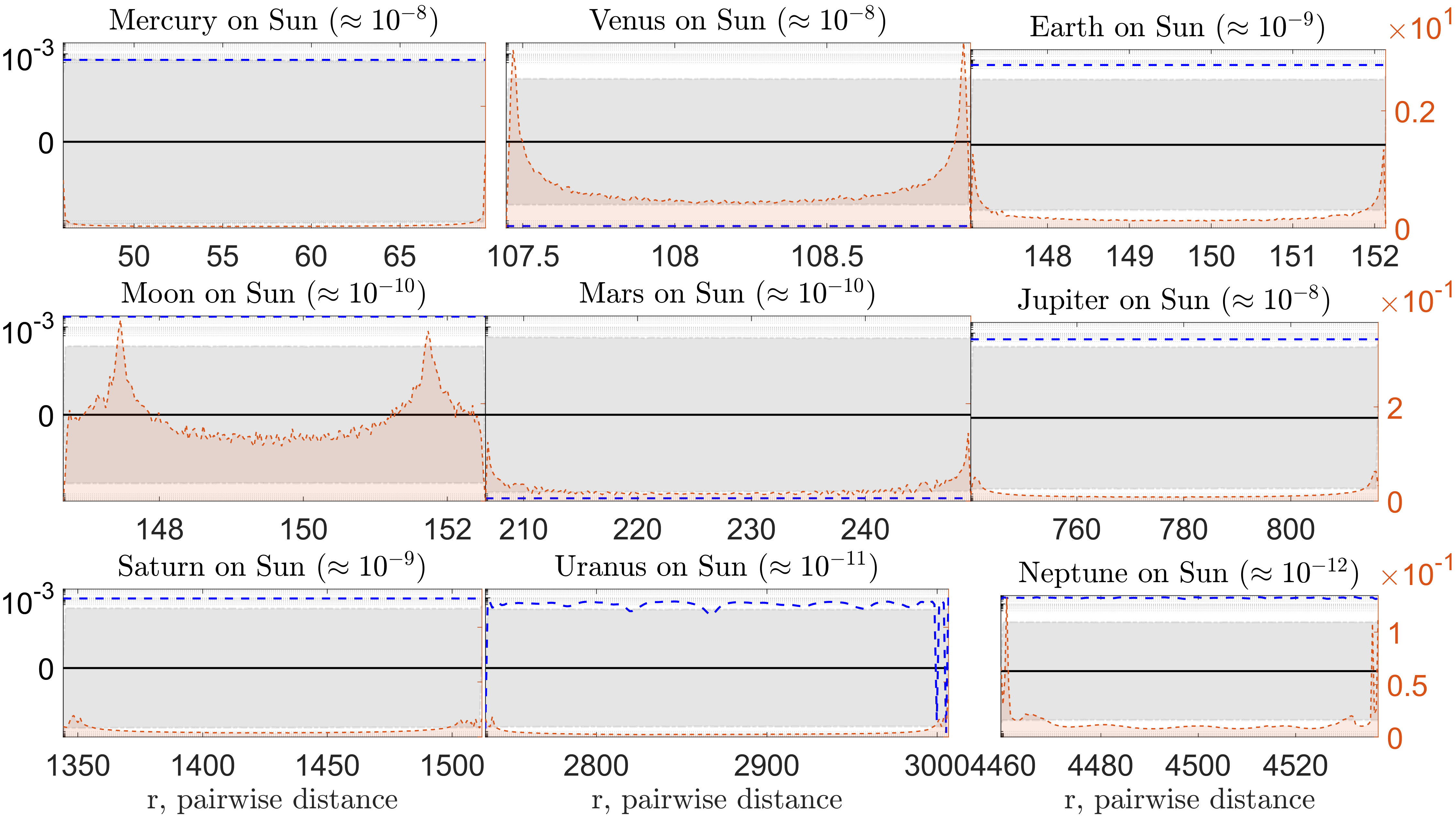}
\end{subfigure}
\caption{Sun-on-planet, $(\lintkernel_{i, 1})_i$, and planet-on-Sun, $(\lintkernel_{1, i})_i$ interaction kernels, compared to those in the Newton and the EIH models (the latter shown as a range, as they depend on other observables besides pairwise distances), shown in terms of relative error w.r.t Newton's gravity, in symmetric-log scale.  Shown in the background is the corresponding distribution of the pairwise distance data used to learn these kernels, i.e. $\rho_{T, i, i'}^{L}$s. We recover the learned kernels at relative error around $10^{-5}$ away from Newton's gravity, and within the range of the EIH range for the Sun-on-planet interactions.  For the planet-on-Sun interactions, the relative errors are around $10^{-4}$ away from Newton's gravity (especially bigger for the two farthest away planets), due to the fact that the strength of the interactions is getting closer to numerical error of around $10^{-10}$ coming from the approximated accelerations.  The absolute scale of the maximum Newton's gravity for each $(i, i')$-pair is shown in the title of corresponding sub-plots.}
\label{fig:JPL_phiEhats_Sun}
\end{figure}

\begin{figure}[H]
\centering
\begin{subfigure}[b]{0.48\textwidth} 
\centering
\includegraphics[width=\textwidth]{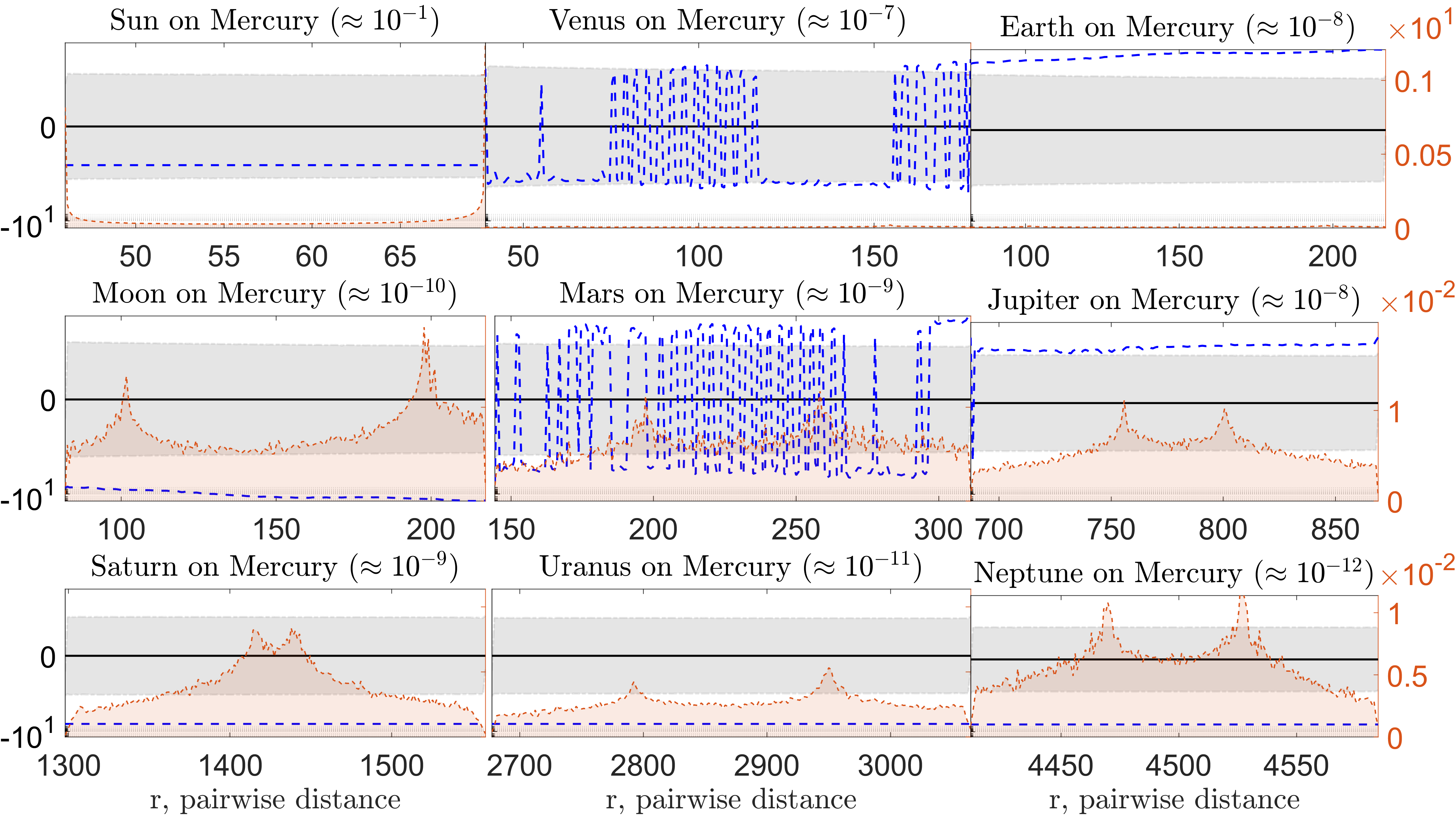} 
\end{subfigure} \\
\begin{subfigure}[b]{0.48\textwidth}
\centering
\includegraphics[width=\textwidth]{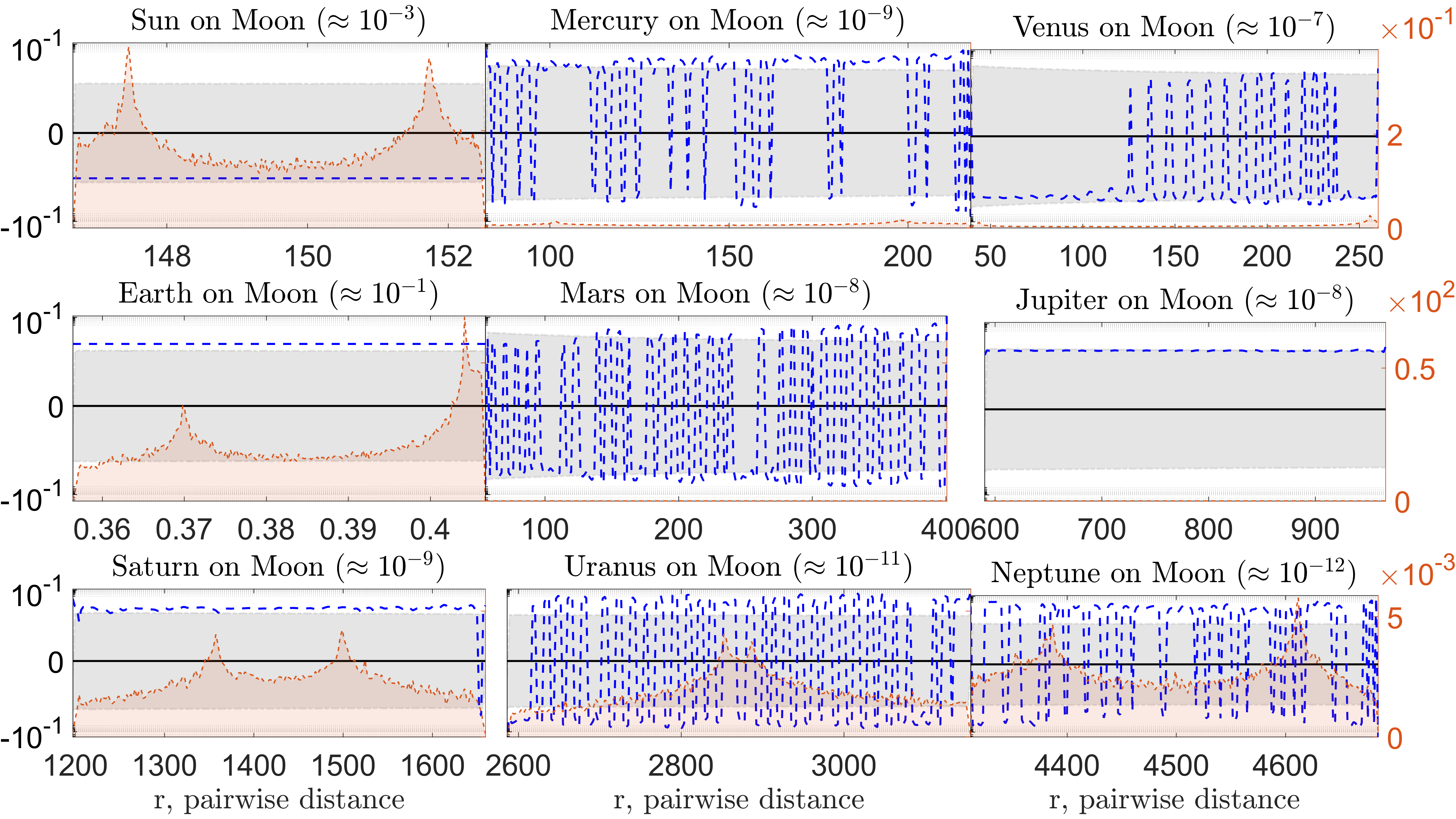}
\end{subfigure} \\
\begin{subfigure}[b]{0.48\textwidth}
\centering
\includegraphics[width=\textwidth]{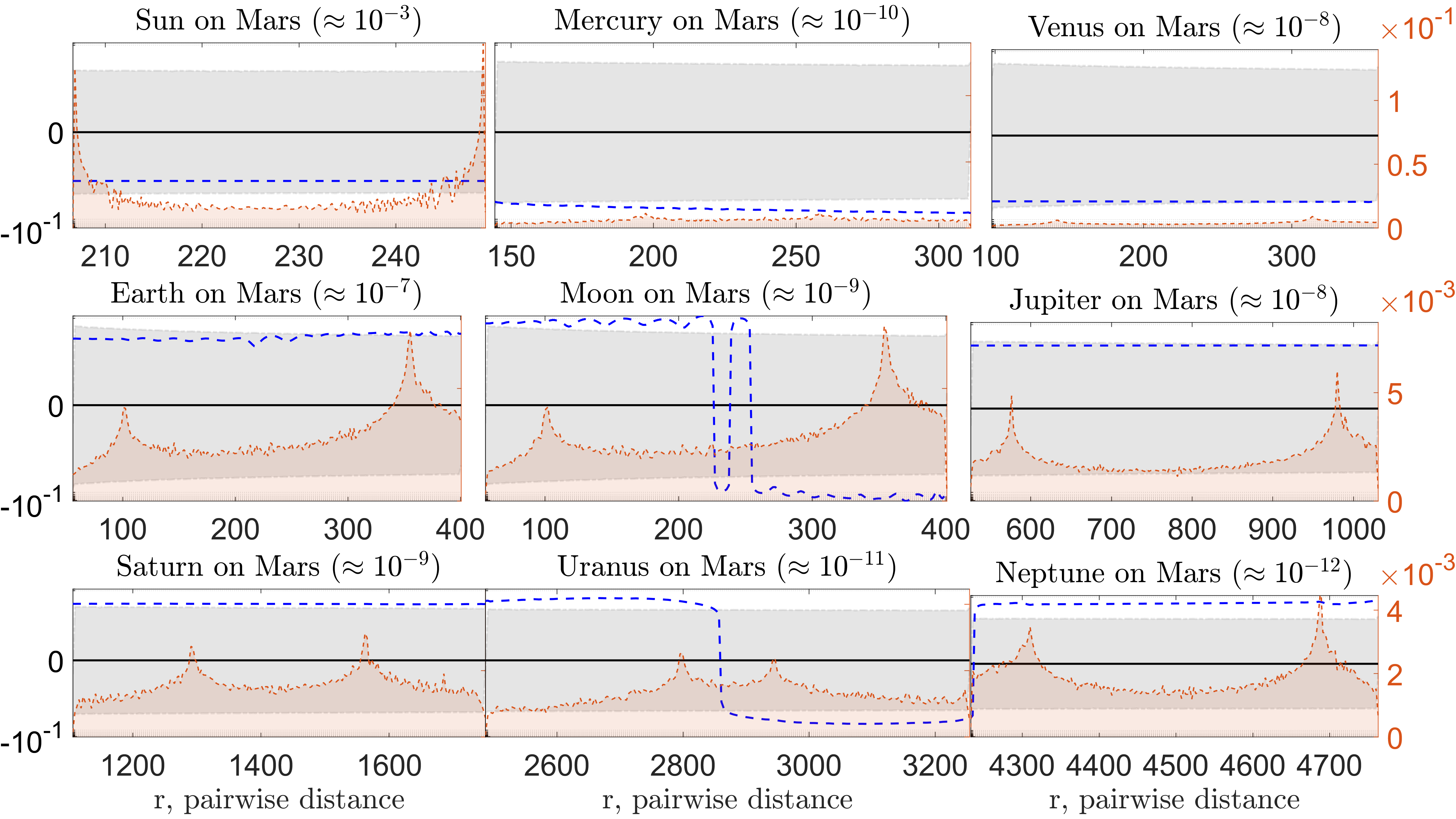}
\end{subfigure}
\caption{Celestial body-on-Mercury ($(\lintkernel_{2, i})_i$), celestial body-on-Moon ($(\lintkernel_{5, i})_i$), and celestial body-on-Mars ($(\lintkernel_{6, i})_i$) interaction kernels vs. Newton's gravity and the EIH range, shown in terms of relative error compared to Newton's gravity in symmetric-log scale.  Shown in the background are the corresponding distribution of pairwise distances, $\rho_{T, i, i'}^{L, }$, used to learn these kernels. As shown in the figures, the learning of interaction kernels on Mars is the easiest; whereas the interaction kernels on Mercury and the Moon both present considerable complications: general relativity effects and lunar effects.  Hence the resulting kernels on Mercury and the Moon show more oscillatory behaviors, especially at small scales.  The absolute scale of the maximum Newton's gravity for each $(i, i')$-pair is shown in the title of the corresponding sub-plots.}
\label{fig:JPL_phiEhats_others}
\end{figure}

\begin{figure}[H]
\centering
\includegraphics[width=0.48\textwidth]{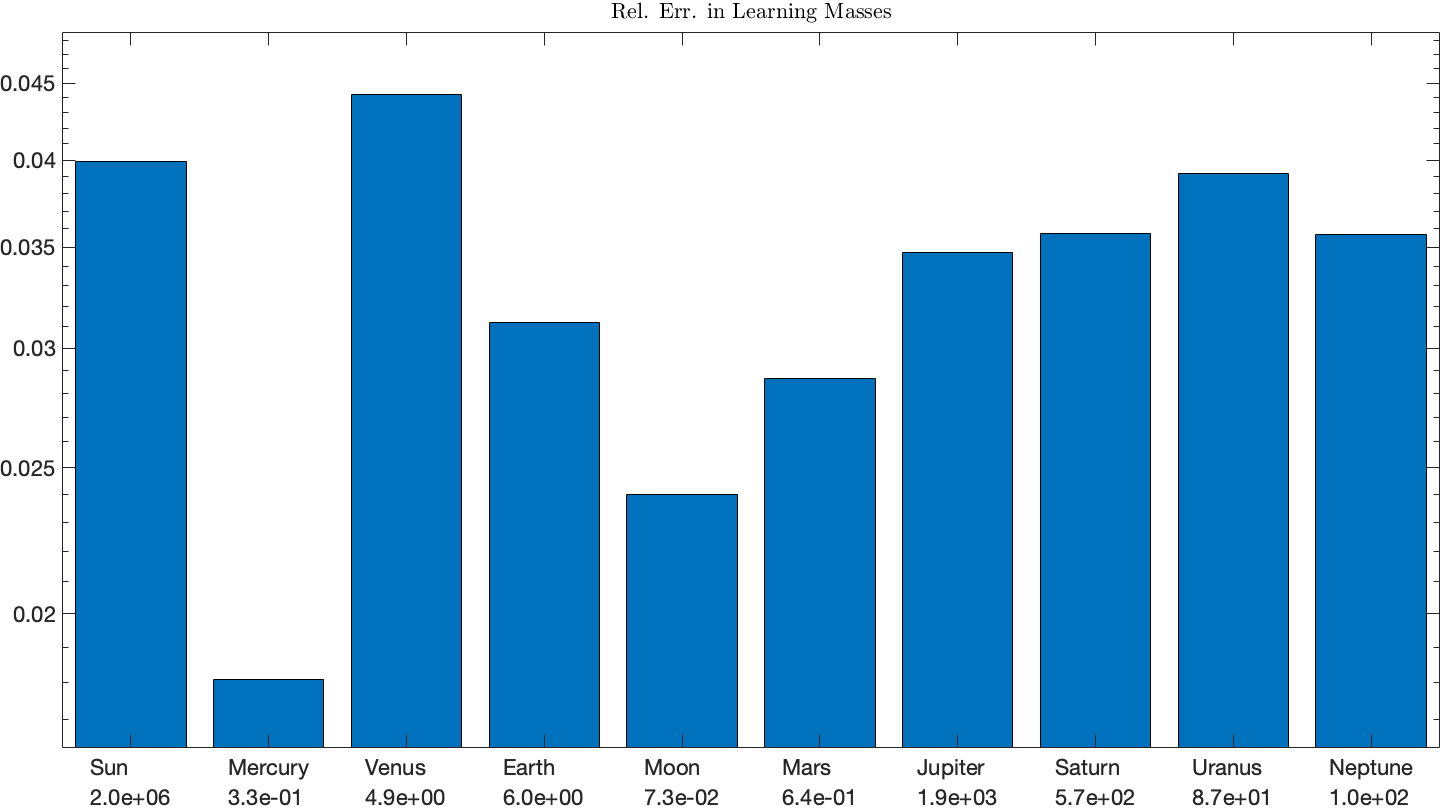} 
\caption{Relative Errors in Learning Masses.  True masses (taken from the JPL database) are shown in black text underneath the names of the corresponding CBs.   Relative errors are shown in blue error bars.  As shown, we have obtained highly accurate results from our de-coupling procedure with two-digit relative accuracy for most of the CBs, notwithstanding the ill-conditioning of the estimation procedure due to the high unevenness of the masses (the Sun takes over $99\%$ of the total mass of the system, and the CBs in the inner Solar system are rather small) and wide-spread distribution of CBs relative distance to the Sun.}
\label{fig:JPL_massErr}
\end{figure}

We take $500$ years of daily position/velocity data ($1500 - 1999$) of the Sun, $8$ major planets, and the Earth's Moon from NASA JPL's DE$430/431$ from their online database (\url{https://ssd.jpl.nasa.gov/horizons.cgi}).  
We perform various learning experiments from subsets of $500$ years of daily data, starting at year $1500$, with the acceleration approximated using a Finite Difference Scheme.  Throughout our experiments, we adopt the following units to conform to the NASA standard: time unit is $t = 1$ day, length unit is $10^{6}$ km, and the unit of mass is $10^{24}$ kg.  The gravitational constant $G$ and speed of light $c$ have been rescaled in these new units (see table $VI$ in SI).  We index the celestial bodies as follows: $1$ is given to the Sun, $2$ to Mercury, $\cdots$, $5$ to Earth, $6$ to the Moon, $\cdots$, lastly $10$ to Neptune.  For the set of learned interaction kernels, $\blintkernel$, we construct the hypothesis spaces using clamped B-splines, with $(S, p) = (90, 4)$.  Here $S$ is the number of sub-intervals in each $[R_{i, i'}^{\min}, R_{i, i'}^{\max}]$, and $p$ is the degree of the clamped B-spline functions.
We present a comparison of our learning results, with training data consisting of the first $450$ years of data, and test/prediction on the subsequent $50$ years, with the observed data, the Newton model and the EIH model \cite{EIH1938} (using the same initial positions/velocities at year $1500$ from the JPL data). This comparison is performed first in terms of learning errors for trajectories, where we achieve higher accuracy than all the other models (see Fig. \ref{fig:JPL_trajErr}), and for the period, aphelion, and perihelion of the trajectories, where we achieve higher accuracy than the other models on most bodies (see Fig. \ref{fig:JPL_PIerr}).  As shown in Figure \ref{fig:JPL_PIerr}, our learned model excels in almost every estimation, except at reproducing the period of Neptune (which might be caused by missing data from Pluto), and reproducing the aphelion/perihelion for Mercury, likely due to the relativity effect not being within our collective dynamics modeling framework.  
Figures \ref{fig:JPL_phiEhats_Sun} and \ref{fig:JPL_phiEhats_others} show the comparison of the learned interaction kernels (Sun-on-planet and planet-on-Sun interaction kernels) versus Newton's gravity, together with a range of general relativity effects according to the EIH model: since these depend on other observables beyond pairwise distance $||\bx_{i'} - \bx_i||$, we represent them as a range over pairwise distance.  The error, $\frac{g_{i, i'} - \text{Newton}_{i, i'}}{\text{Newton}_{i, i'}}$, is shown in symmetric-log scale with $\text{Newton}_{i, i'} = \frac{Gm_{i'}}{r^3}$ and $f_{g, i'} = \lintkernel_{i, i'}$ or $\text{EIH}^{\max \, \text{or} \, \min}_{i, i'}$.
As shown in Figure \ref{fig:JPL_phiEhats_others}, we are able to recover a set of learned interaction kernels, which oscillate within the EIH range around Newton's gravity.  The computation of the EIH range is detailed in the section ``Learning Results'' in SI.  We are also interested in how our learned model recreates the perihelion precession rate of the orbits of Mars, Mercury, and the Moon, and we compare it to the precession rates calculated from the JPL data, and from dynamical data generated by the Newton and the EIH models, see Table \ref{tab:PR}. Finally, we learn the mass of each CB in the Solar System from an improved de-coupling procedure first introduced in \cite{ZMM2020}.  The result is shown in figure \ref{fig:JPL_massErr}, and details of the procedure can be found in the section ``Estimating Masses of Celestial Bodies'' in SI.
\noindent\textbf{Observation:} The Sun-on-planet interaction kernels $(\lintkernel_{i, 1})_i$, are the main driving forces behind the celestial motions in the Solar System, due to their massive relative scales.  We are able to recover these kernels in a form between Newton's gravity and the EIH range, with point-wise relative errors at the scale around $10^{-5}$.  A similar observation can be also made on other interaction kernels; however due to the substantially smaller relative scales of other interaction kernels, numerical approximation errors (especially the ones from approximating accelerations) and learning errors, it results in oscillatory behaviors of our learned kernels.  We also want to point out that the learned kernels contain partial information about general relativity, resulting in improved prediction of perihelion precession rates when compared to the Newton model.
\section{Conclusion}\label{sec:conclude}
We have demonstrated the effectiveness of our learning methods to study the celestial motion of the Solar System using NASA JPL's development ephemerides.  Our learned model offers better performance than the Newton and EIH models in terms of trajectory prediction, and recreates trajectories which also preserve various geometric properties, such as the period, aphelion, and perihelion, with high accuracy.  
 
Furthermore, our learning methods can recreate dynamics which given perihelion precession rates for the orbits of Mars, Mercury, and the Moon that are closer to the observed rates than the ones produced by the Newton model.  The dynamics from our learned model can also be used to learn the mass of each celestial body.  Aided by geometric machine learning techniques, such as the approach discussed in \cite{MMQZ2021}, our learning method can be extended to study galaxy dynamics or the Solar System from a relativistic point of view.
\begin{acknowledgments}
\textbf{Acknowledgment}: MM and MZ designed the research; JM and MZ obtained and processed the observation data; JM and MZ developed algorithms; JM and MZ analyzed the numerical results; all authors wrote the manuscript.
MM is grateful for partial support from NSF-1837991, NSF-1913243, FA9550-20-1-0288, and to the Simons Foundation for the Simons Fellowship for the year '20-'21. Prisma Analytics, Inc. provided computing equipment and support.  Software package (including data) can be found on our GitHub page (\url{https://github.com/MingZhongCodes/LearningDynamics/}).
\end{acknowledgments}
\appendix
\section{Celestial Dynamics Models}
We describe in this section the equations of motion for various celestial dynamics models: the Jet Propulsion Laboratory model (JPL), the Einstein-Infeld-Hoffmann model (EIH), and the Newton model (Newton).
\subsection{The JPL Model}
In the JPL model, the set of contributing celestial bodies (CBs) is comprised of $N = 11$ CBs:  Sun, 8 major planets, Moon of Earth, and Pluto.  
The equations of motion are
\begin{equation}\label{eq:JPL_eq}
\begin{aligned}
\dot\bx_i &= \bv_i \\
\dot\bv_i &= \sum_{\substack{i' = 1 \\ i' \neq i}}^{N}\frac{Gm_{i'}}{\norm{\bx_{i'} - \bx_{i}}^3}\cdot(\bx_{i'} - \bx_i)\Bigg\{1 - \frac{4}{c^2}\sum_{\substack{i'' = 1 \\ i'' \neq i}}^{N}\frac{Gm_{i''}}{\norm{\bx_{i''} - \bx_i}} -\frac{1}{c^2}\sum_{\substack{i'' = 1 \\ i'' \neq i'}}^{N}\frac{Gm_{i''}}{\norm{\bx_{i''} - \bx_{i'}}} \nonumber\\
  &\quad + \frac{2\norm{\bv_{i'} - \bv_i}^2 - \norm{\bv_i}^2}{c^2} - \frac{3}{2c^2}\big(\dotp{\frac{\bx_{i'} - \bx_{i}}{\norm{\bx_i - \bx_{i'}}}, \bv_{i'}}\big)^2 + \frac{1}{2c^2}\dotp{\bx_{i'} - \bx_{i}, \dot\bv_{i'}}\Bigg\} \nonumber \\
  &\quad - \sum_{\substack{i' = 1 \\ i' \neq i}}^{N}\frac{Gm_{i'}}{\norm{\bx_{i'} - \bx_{i}}^3}\dotp{\bx_{i'} - \bx_i, \frac{3\bv_{i'} - 4\bv_i}{c^2}}\cdot(\bv_{i'} - \bv_{i}) \nonumber \\
  &\quad + \frac{7}{2c^2}\sum_{\substack{i' = 1 \\ i' \neq i}}^{N}\frac{Gm_{i'}}{\norm{\bx_{i'} - \bx_{i}}}\dot\bv_{i'} + \sum_{i' = N + 1}^{N + 3}\frac{Gm_{i'}}{\norm{\bx_{i'} - \bx_i}}(\bx_{i'} - \bx_i) + \sum_{\text{$297$ asteroids}}\mbf{F},
\end{aligned}
\end{equation}
for $i = 1,  \ldots, N$.   Here $G$ is the gravitational constant, $m_i$ is the mass of the $i^{th}$ CB, $c$ is the speed of light in vacuum, and $\bx_i$, $\bv_i$ represent the position and velocity of the barycenter of the $i^{th}$ CB.  The additional $3$ CBs appearing the next-to-last sum are Ceres, Pallas, and Vesta, which are only used for the calculation of $\dot\bv_i$ for the first $N$ CBs.  The last term gives the forces from a set of $297$ asteroids; these forces are considered only for perturbations on the Earth, Moon and Mars.  The motions of Ceres, Pallas, Vesta and $297$ other asteroids are given special treatment, see the details outlined in \cite{SW2007}. Moreover, the JPL model also considers other possible physical laws, in particular those included in the Lunar Theory, to make the evolution of the celestial motion as close to the true observation data as possible.  For details, see \cite{SW2007}.
\subsection{The Einstein-Infeld-Hoffmann Model}\label{sec:EIH_eq}
The Einstein-Infeld-Hoffmann model (EIH) uses the equations of motion based on a first-order expansion of Einstein's field equations of general relativity.  Given a system of $N$ CBs, indexed by table \ref{tab:CB_index} (and distinct from that considered in equations for the JPL model), the barycentric acceleration vector of the $i^{th}$ CB is given by 
\begin{equation}\label{eq:EIH_eq}
\begin{aligned}
\dot\bx_i &= \bv_i, \\
\dot\bv_i &= \sum_{\substack{i' = 1 \\ i' \neq i}}^{N}\frac{Gm_{i'}}{\norm{\bx_{i'} - \bx_{i}}^3}\cdot(\bx_{i'} - \bx_i)\Bigg\{1 - \frac{4}{c^2}\sum_{\substack{i'' = 1 \\ i'' \neq i}}^{N}\frac{Gm_{i''}}{\norm{\bx_{i''} - \bx_i}} -\frac{1}{c^2}\sum_{\substack{i'' = 1 \\ i'' \neq i'}}^{N}\frac{Gm_{i''}}{\norm{\bx_{i''} - \bx_{i'}}} \nonumber\\
  &\quad + \frac{2\norm{\bv_{i'} - \bv_i}^2 - \norm{\bv_i}^2}{c^2} - \frac{3}{2c^2}\big(\dotp{\frac{\bx_{i'} - \bx_{i}}{\norm{\bx_i - \bx_{i'}}}, \bv_{i'}}\big)^2 + \frac{1}{2c^2}\dotp{\bx_{i'} - \bx_{i}, \dot\bv_{i'}}\Bigg\} \nonumber \\
  &\quad - \sum_{\substack{i' = 1 \\ i' \neq i}}^{N}\frac{Gm_{i'}}{\norm{\bx_{i'} - \bx_{i}}^3}\dotp{\bx_{i'} - \bx_i, \frac{3\bv_{i'} - 4\bv_i}{c^2}}\cdot(\bv_{i'} - \bv_{i})
 + \frac{7}{2c^2}\sum_{\substack{i' = 1 \\ i' \neq i}}^{N}\frac{Gm_{i'}}{\norm{\bx_{i'} - \bx_{i}}}\dot\bv_{i'},
\end{aligned}
\end{equation}
for $i = 1, \ldots, N$.
\subsection{The Newton Model}
The Newton model uses the celebrated Newton's universal law of gravitation as its equations of motion,
\begin{equation}\label{eq:Newton}
\begin{aligned}
\dot\bx_i &= \bv_i, \\
\dot\bv_i &= \sum_{i' = 1, i' \neq i}^N \frac{Gm_{i'}}{\norm{\bx_{i'} - \bx_i}^3}(\bx_{i'} - \bx_i), 
\end{aligned}\quad i = 1, \ldots, N.
\end{equation}
\section{Learning Framework}\label{sec:framework_SI}
We describe in this section the framework used to find the set of learned interaction kernels, by minimizing the error functional
\begin{equation}\label{eq:loss_func}
\mE_{L}(\bintkernelvar) = \frac{1}{LN}\sum_{l, i = 1}^{L, N}\norm{\ddot\bx_i - \sum_{i' = 1}^N\intkernelvar_{i, i'}(\norm{\bx_{i'}(t_l) - \bx_i(t_l)})(\bx_{i'}(t_l) - \bx_i(t_l))}^2.
\end{equation}
over $\bintkernelvar = [\intkernelvar_{i, i'}]_{i, i' = 1}^N$ with each $\intkernelvar$ in a suitable hypothesis space of functions $\hypspace_{i, i'}$, with $\dim(\hypspace_{i, i'}) = n_{i, i'}$. The loss functional leads to a natural dynamics-adapted probability measure, 
\begin{equation}\label{eq:rho}
\rho_{T, i, i'}^L = \frac{1}{L}\sum_{l = 1}^L\delta_{r_{i, i'}(t_l)}(r),
\end{equation}
where $r_{i, i'}(t_l) = \norm{\bx_{i'}(t_l) - \bx_i(t_l)}$, and $\delta$ is the Dirac measure. The measure $\rho_{T, i, i'}^L$ measures the time-averaged appearance of observed pairwise distance data along trajectories.
\subsection{Related Works}
In \cite{BFHM17}, a variational approach was introduced for learning the interaction kernel from observations of first order homogeneous particle systems; and convergence properties were analyzed when $N$, the number of particles, goes to infinity -- namely the mean field limit.  We extended this learning approach in \cite{LZTM2019} to heterogeneous particle systems of first and second order; and studied the convergence in $M$, the number of different initial conditions, for fixed $N$.  In \cite{ZMM2020} we discussed the steady state behavior of the dynamics evolved using our learned models, in particular for dynamical systems displaying emergent behaviors. An extended and complete learning theory for a broad class of second-order models is analyzed and investigated in \cite{MTZM2020}.   A learning theory on first-order dynamics constrained on Riemannian manifolds is analyzed and investigated in \cite{MMQZ2021}.
\subsection{Numerical Algorithms}\label{sec:algorithms}
We discuss in detail the algorithm used to solve \eqref{eq:loss_func}, over the set of finite dimensional hypothesis spaces $\{\hypspace_{i, i'}\}_{i, i' = 1}^N$.  First, when $i \neq i'$, we take 
\begin{align*}
R_{i, i'}^{\min} &\coloneqq \{\min\}_{l = 1}^L \norm{\bx_{i'}(t_l) - \bx_i(t_l)}, \\
R_{i, i'}^{\max} &\coloneqq \{\max\}_{l = 1}^L \norm{\bx_{i'}(t_l) - \bx_i(t_l)};
\end{align*}
when $i = i'$, we take $R_{i, i}^{\min,\max} \coloneqq 0,1$ respectively.  Next, we use Clamped B-spline functions\footnote{Other kinds of basis functions can also be used, see the examples in \cite{LZTM2019}.} of degree $p_{i, i'} = p \ge 2$ so that the estimated interaction kernels would at least have continuous first derivatives; furthermore, we are using a uniform $p$ for all $i \neq i'$.  Then these B-spline functions are used as the basis functions and build $\hypspace_{i, i'}$ over a uniform partition of $[R_{i, i'}^{\min}, R_{i, i'}^{\max}]$ with the number of sub-intervals equaling $S_{i, i'} = S$\footnote{We use a uniform $S$ here to simplify the discussion; in practice, a non-uniform series $S_{i, i'}$s actually helps in reducing the computational time.} (note that $n_{i, i'} = n = p + S$ for the Clamped B-spline basis), when $i \neq i'$; when $i = i'$, we simply take $p_{i, i} = 0$ and $S_{i, i} = 1$, hence $n_{i, i} = 1$.  Next, for $i = 1, \ldots, N$, we assemble the basis matrices $\Psi^{l}_i \in \R^{d \times ((N - 1)n + 1)}$, and the right hand size vector $\vec{d}^{l}_i \in \R^{d}$ in the following way.  For $\eta_i = 1, \ldots, (N - 1)n + 1$, 
\[
\Psi^l_i(:, \eta_i) = \basis_{i, i', \eta_{i, i'}}(r_{i, i'}(t_l))\br_{i, i'}(t_l).
\]
Here $\eta_{i, i'}$ is computed as follows: when $1 \le \eta_i \le n$ (when $i \neq 1$) or $1$ (when $i = 1$), we take $\eta_{i, i'} = \eta_i$; when $\eta_i > n$ (when $i > 1$) or $1$ (when $i = 1$), we find the $i_*$ such that $\sum_{i'' = 1}^{i_*} n_{i, i''} < \eta_i < \sum_{i'' = 1}^{i_* + 1}n_{i, i''}$ (recall that $n_{i, i''} = n$ when $i \neq i''$ and $n_{i, i''} = 1$ when $i = i''$), then let $i' = i_* + 1$ and set $\eta_{i, i'} = \eta_i - \sum_{i'' = 1}^{i_*} n_{i, i''}$.  For $\vec{d}^{l}_i$, we simply perform the assignment, i.e. set $\vec{d}^{l}_i = \dot\bv_i(t_l)$.  We then assemble $A^L_i \in \R^{n_i \times n_i}$ and $\vec{b}_i^L \in \R^{n_i \times 1}$ as follows
\[
A^L_i = \frac{1}{L}\sum_{l = 1}^L(\Psi^l_i)^\top\Psi^l_i \quad \text{and} \quad \vec{b}^L_i = \frac{1}{L}\sum_{l = 1}^L(\Psi^l_i)^\top\vec{d}^l_i.
\]
We solve for $\widehat{\vec{\alpha}}_i$ from the system $A^L_i\vec{\alpha} = \vec{b}^L_i$ using a pseudoinverse, and assemble $\lintkernel_{i, i'} \coloneqq \sum_{\eta_{i, i'} = 1}^{n_{i, i'}} \hat{\alpha}_{i, i', \eta_{i, i'}}\basis_{i, i', \eta_{i, i'}}$, here $\basis_{i, i', \eta_{i, i'}}$ is the $\eta_{i, i'}^{th}$ basis function for $\hypspace_{i, i'}^{1D}$.  The actual implementation can be easily parallelized in $l$, see similar implementations done in \cite{LZTM2019, ZMM2020, MTZM2020, MMQZ2021}.
\subsection{Computational Complexity}
We present a detailed discussion on the total computational complexity of solving the inference problem for learning the unknown interaction kernels.  In order to compute the individual learning interval, i.e. $[R_{i, i'}^{\min}, R_{i, i'}^{\max}]$ for $i, i' = 1, \cdots, N$, we need to perform $\mO(N^2)$ computation of pairwise distances at each time instance, hence ending up with a total of $\mO(LN^2)$ for computing pairwise distances.  Then in assembling $\Psi^l_i$ for each celestial body, the algorithm does $\mO(Nn)$ basis evaluations at each time step and for each celestial body, thus ending up costing a total of $\mO(LN^2n)$ basis evaluations.  The assembly of $\vec{d}^{l}_i$ is based on value assignments hence it is negligible.  Then for the assembly of $A^L_i$, it needs to perform a total of $\mO(LdN^2n^2)$ operations for computing $(\Psi^l_i)^\top\Psi^l_i$ for each celestial body, hence we need to do a total of $\mO(LdN^3n^2)$ operations for assembling $A^L_i$.  Similarly, we need to perform a total of $\mO(LdN^2n)$ operations for the assembly of the $N$ $\vec{b}^L_i$s.  Finally, in solving the linear systems, it does a total of $N^4n^3$ operations.  Therefore, the total computing time needed for the whole learning problem is 
\[
T_{\text{tol}} = LN^2 + LN^2n + LdN^3n^2 + LdN^2n^2 + N^4n^3
\]
since we have $Ld \gg Nn$, i.e. we are processing hundreds of years of observation data, we have $T_{\text{tol}} \approx LdN^3n^2$; the computational bottleneck is caused by the assembly of the learning matrix $A^L_i$s.  In the case of learning from $500$ years of position/velocity data (i.e. $L \approx 500 * 365$, $N = 10$, $d = 3$, and $n \approx 100$), the assembly of $A^L_i$s would require a total of $4.445 \cdot 10^{12}$ operations.  Due to the extremely large size of $L$, we need to perform a parallelization of the learning algorithm in $L$ by splitting the one long system trajectory into many short trajectories.

As far as memory is concerned, it takes $\mO(LdN)$ to store $500$ years of observed position, velocity data of $N = 10$ celestial bodies, amounting to roughly $1.095 \cdot 10^{7}$ data points, which is considered possible for modern day workstations.  However, in the assembly of the $A^L_i$s, one needs to construct all $\Psi^l_i$s, leading to a total of $LdN^2n \approx 4.933 \cdot 10^9$ data points, pushing into the domain of supercomputers.  One remedy would be to process the $\Psi^l_i$s in sequential order, however this would be very time consuming.  The proper choice is to perform the assembly in parallel in $L$, thus significantly reducing the time needed to perform the assembly in terms of both memory and computing time, and making the learning algorithm able to be run on personal laptops.
\subsection{Approximating Acceleration Data}
Next we discuss how the derivative should be approximated based on the limited precision for double-floating points.  To simplify the discussion, we will start from the a set of scalar function values, $\{f(t_l)\}_{l = 1}^L$, and assume we are able to obtain a limited amount of finer grid points of $f$, i.e. $f(t_l + h)$s with $h \ll t_{l + 1} - t_l$.  In the JPL data case, we have 
\[
t_{l + 1} - t_l = 1 \quad \text{and} \quad h = \frac{1}{24 * 60} \approx 6.9444 \cdot 10^{-4}.
\]
We consider a series of $(2k + 1)$-point central schemes, for being as accurate as other $(2k + 1)$-point schemes but using one point fewer, which uses the following formula
\[
f'(t_l) \approx \mathcal{C}^{2k + 1}_h(f(t_l)) = \frac{\sum_{i = 1}^k a_i(f(t_l + ih) - f(t_l - ih))}{b_{2k + 1}h},
\]
where $\mathcal{C}^{2k + 1}_h$ is the $(2k + 1)$-point central scheme operator.  Here $a_i$s are integers, and $b_{2k + 1}$ is some positive even integer multiple of $k$.  Note that $f(t_l)$ is not used in the formula, but since the lattice covers $f(t_l)$, we will still call it a $(2k + 1)$-point scheme; see the following table.
\begin{table}[H]
\centering
\begin{tabular}{c | c | c}
\hline
$N$ & $N$-point Stencil                       & Leading Term \\
\hline
$3$ & $a_1 = 1, b_3 = 2$                      & $\frac{f'''(t_l)h^2}{6}$ \\
\hline
$5$ & $a_is = [8, -1], b_5 = 12$              & $-\frac{f^{(5)}(t_l)h^4}{30}$\\
\hline
$7$ & $a_is = [45, -9, 1], b_7 = 60$          & $\frac{f^{(7)}(t_l)h^6}{420}$ \\
\hline
\end{tabular}  
\caption{Various Central Schemes}
\label{tab:CS_points} 
\end{table}
Hence a $(2k + 1)$-point central scheme can give a local error at the order of $h^{2k}$.  Now consider the limited precision double-floating double representation, we can assume that
\[
f_l^{\epsilon_l} = f^{\epsilon_l}(t_l) = f(t_l)(1 + \epsilon_l).
\]
Here $\abs{\epsilon_l} \le 10^{-16}$.  Then
\[
\begin{aligned}
\mathcal{C}^{2k + 1}_h(f_l^{\epsilon_l}) &= \mathcal{C}^{2k + 1}_h(f(t_l)(1 + \epsilon(t_l))) \\
&= \frac{\sum_{i = 1}^k a_i(f(t_l + ih)(1 + \epsilon(t_l + ih)) - f(t_l - ih)(1 + \epsilon(t_l - ih)))}{b_{2k + 1}h}\\
&\approx  \mathcal{C}^{2k + 1}_h(f_l) + \frac{\sum_{i = 1}^k a_i(f(t_l + ih)\epsilon(t_l + ih) - f(t_l - ih)\epsilon(t_l - ih))}{b_{2k + 1}h}.
\end{aligned}
\]
Here we are assuming that
$
f(t_l - kh) \approx \cdots \approx f(t_l - h) \approx f(t_l) \approx f(t_l + h) \approx \cdots \approx f(t_l + kh),
$
since $h$ is very small, and $f$ at $t_l$ is continuous; hence we have
\[
\begin{aligned}
&\abs{\frac{\sum_{i = 1}^k a_i(f(t_l + ih)\epsilon(t_l + ih) - f(t_l - ih)\epsilon(t_l - ih))}{b_{2k + 1}h}} \\
&\approx \abs{f(t_l)}\abs{\frac{\sum_{i = 1}^k a_i(\epsilon(t_l + ih) - \epsilon(t_l - ih))}{b_{2k + 1}h}} \le \frac{2\sum_{i = 1}^k\abs{a_i}}{b_{2k + 1}h}\abs{f_l}\cdot 10^{-16}
\end{aligned}
\]
Requiring the balance
\[
\abs{c_{2k + 1}h^{2k}} \approx \frac{2\sum_{i = 1}^k\abs{a_i}}{b_{2k + 1}h}\abs{f_l}\cdot 10^{-16},
\]
we obtain
\[
h \approx \sqrt[2k + 1]{\frac{2\sum_{i = 1}^k\abs{a_i}}{b_{2k + 1}} \cdot \frac{\abs{f_l}}{\abs{c_{2k + 1}}} \cdot 10^{-16}}.
\]
As $k \rightarrow \infty$, $h \rightarrow 1$ (assuming $\frac{\sum_{i = 1}^k\abs{a_i}}{b_{2k + 1}\abs{c_{2k _ 1}}} \rightarrow $ const.), hence higher order schemes are not desired; unless as $k \rightarrow \infty$, $\frac{\sum_{i = 1}^k\abs{a_i}}{b_{2k + 1}\abs{c_{2k _ 1}}} \rightarrow \infty$, then it might make sense to use high order schemes.  Another restriction is that $f$ is required to be smooth enough for us to be able take at many derivatives as possible.  Let us focus on the $3$, $5$, and $7$-pt schemes for now, a similar derivation can be applied to $9$-pt and beyond.  In order to find the optimal scheme, we need to know $\abs{f(t_l)}$ as well as $\abs{c_{2k + 1}}$ (which just corresponds to higher order derivatives of $f(t_l)$).  We drive the following errors bounds based on the averaged speed of each celestial body in table \ref{tab:EUB_CSS}.
\begin{table}[H]
\centering
\begin{tabular}{c || c  | c | c}
\hline
Celes. Body              & $\abs{f(t_l)}$            & $\abs{c_{2k + 1}}$             & Err. UB\\
\hline
\multirow{3}{*}{Mercury} & \multirow{3}{*}{$4.1386$} & $3$-pt: $8.5144 \cdot 10^{-2}$ & $4.1062 \cdot 10^{-8}$\\
                         &                           & $5$-pt: $1.4610 \cdot 10^{-1}$ & $9.2838 \cdot 10^{-13}$\\
                         &                           & $7$-pt: $2.8681 \cdot 10^{-1}$ & $1.0926 \cdot 10^{-12}$\\
\hline
\multirow{3}{*}{Venus}   & \multirow{3}{*}{$3.0240$} & $3$-pt: $1.6271 \cdot 10^{-3}$ & $7.8509 \cdot 10^{-10}$\\
                         &                           & $5$-pt: $1.9414 \cdot 10^{-4}$ & $6.5323 \cdot 10^{-13}$\\
                         &                           & $7$-pt: $6.1392 \cdot 10^{-5}$ & $7.9834 \cdot 10^{-13}$\\
\hline
\multirow{3}{*}{Earth}   & \multirow{3}{*}{$2.5747$} & $3$-pt: $3.3395 \cdot 10^{-4}$ & $1.6142 \cdot 10^{-10}$\\
                         &                           & $5$-pt: $1.4366 \cdot 10^{-5}$ & $5.5614 \cdot 10^{-13}$\\
                         &                           & $7$-pt: $1.8822 \cdot 10^{-6}$ & $6.7972 \cdot 10^{-13}$\\
\hline 
\multirow{3}{*}{Moon}    & \multirow{3}{*}{$2.6630$} & $3$-pt: $3.7006 \cdot 10^{-4}$ & $1.7885 \cdot 10^{-10}$\\
                         &                           & $5$-pt: $1.6669 \cdot 10^{-5}$ & $5.7521 \cdot 10^{-13}$\\
                         &                           & $7$-pt: $2.3629 \cdot 10^{-6}$ & $7.0303 \cdot 10^{-13}$\\
\hline
\multirow{3}{*}{Mars}    & \multirow{3}{*}{$2.0822$} & $3$-pt: $5.5023 \cdot 10^{-5}$ & $2.6835 \cdot 10^{-11}$\\
                         &                           & $5$-pt: $7.1423 \cdot 10^{-7}$ & $4.4976 \cdot 10^{-13}$\\
                         &                           & $7$-pt: $3.4283 \cdot 10^{-8}$ & $5.4970 \cdot 10^{-13}$\\
\hline
\multirow{3}{*}{Jupiter} & \multirow{3}{*}{$1.1318$} & $3$-pt: $9.4759 \cdot 10^{-8}$  & $2.0868 \cdot 10^{-13}$\\
                         &                           & $5$-pt: $5.7075 \cdot 10^{-12}$ & $2.4447 \cdot 10^{-13}$\\
                         &                           & $7$-pt: $2.9503 \cdot 10^{-14}$ & $2.9880 \cdot 10^{-13}$\\
\hline
\multirow{3}{*}{Saturn}  & \multirow{3}{*}{$0.8381$} & $3$-pt: $5.3127 \cdot 10^{-9}$  & $1.2325 \cdot 10^{-13}$\\
                         &                           & $5$-pt: $4.5797 \cdot 10^{-13}$ & $1.8103 \cdot 10^{-13}$\\
                         &                           & $7$-pt: $5.1516 \cdot 10^{-17}$ & $2.2126 \cdot 10^{-13}$\\
\hline
\multirow{3}{*}{Uranus}  & \multirow{3}{*}{$0.5875$} & $3$-pt: $1.4955 \cdot 10^{-10}$ & $8.4672 \cdot 10^{-14}$\\
                         &                           & $5$-pt: $4.4185 \cdot 10^{-15}$ & $1.2690 \cdot 10^{-13}$\\
                         &                           & $7$-pt: $2.0032 \cdot 10^{-17}$ & $1.5510 \cdot 10^{-13}$\\
\hline
\multirow{3}{*}{Neptune} & \multirow{3}{*}{$0.4666$} & $3$-pt: $1.3052 \cdot 10^{-11}$ & $6.7197 \cdot 10^{-14}$\\
                         &                           & $5$-pt: $1.4788 \cdot 10^{-16}$ & $1.0079 \cdot 10^{-13}$\\
                         &                           & $7$-pt: $8.4746 \cdot 10^{-23}$ & $1.2318 \cdot 10^{-13}$\\
\hline              
\end{tabular}  
\caption{Upper Bound for the errors at approximating derivatives using various central schemes}
\label{tab:EUB_CSS} 
\end{table}
Based on our analysis of the acceleration accuracies, for the celestial bodies in the inner Solar system, we will use the $5$-point scheme; and for the celestial bodies in the outer Solar system, we will use the $3$-point scheme.   Since the average speed of the Sun is $1.0911 \cdot 10^{-3}$, we have shown in table \ref{tab:Sun_CSS} that the $5$-point scheme is the best.
\begin{table}[H]
\centering
\begin{tabular}{c || c  | c | c}
\hline
Celes. Body          & $\abs{f(t_l)}$                          & $\abs{c_{2k + 1}}$             & Err. UB\\
\hline
\multirow{3}{*}{Sun} & \multirow{3}{*}{$1.0911 \cdot 10^{-3}$} & $3$-pt: $8.6933 \cdot 10^{-9}$ & $4.3495 \cdot 10^{-15}$\\
                     &                                         & $5$-pt: $2.1434 \cdot 10^{-8}$ & $2.3568 \cdot 10^{-16}$\\
                     &                                         & $7$-pt: $1.0423 \cdot 10^{-7}$ & $2.8805 \cdot 10^{-16}$\\
\hline              
\end{tabular}  
\caption{Upper Bound for the errors at approximating derivatives using various central schemes for Sun}
\label{tab:Sun_CSS} 
\end{table}
\subsection{Symplectic Integration}
In order to preserve the Lagrangian and the Hamiltonian system associated to the Newton model, as well as the dynamical systems driven by our interaction kernel estimators, we use a forth order Leapfrog scheme to handle the long time integration and obtain sufficiently stable trajectories.  The usual second order Leapfrog scheme tries to integrate the following ODE, $\ddot\bx = F(\bx)$, for $t \in [0, T]$, where $\bx$ is the position data, with initial conditions $\bx(0) = \bx_0$ and $\dot\bx(0) = \bv(0)$.  Then the second order Leapfrog evolves the ODE from $t_l$ to $t_{l + 1}$ with the following update scheme
\[
\begin{aligned}
\bx_{t_{l + 1}} &= \bx_{t_l} + \bv_{t_l}\Delta t + \frac{1}{2}F(\bx_{t_l})\Delta t^2, & \bv_{t_{l + 1}} &= \bv_{t_l} + \frac{1}{2}(F(\bx_{t_l}) + F(\bx_{t_l + 1}))\Delta t.
\end{aligned}
\]
When combined with the Yoshida algorithm, one can obtain a forth order Leapfrog in the following way
\[
\begin{aligned}
\bx_{t_l}^1 &= \bx_{t_l} + c_1\bv_{t_l}\Delta t,& \bv_{t_l}^1 &= \bv_{t_l} + d_1F(\bx_{t_l}^1)\Delta t, \\
\bx_{t_l}^2 &= \bx_{t_l}^1 + c_2\bv_{t_l}^1\Delta t, & \bv_{t_l}^2 &= \bv_{t_l}^1 + d_2F(\bx_{t_l}^2)\Delta t, \\
\bx_{t_l}^3 &= \bx_{t_l}^2 + c_3\bv_{t_l}^2\Delta t, & \bv_{t_l}^3 &= \bv_{t_l}^2 + d_3F(\bx_{t_l}^3)\Delta t, \\
\bx_{t_{l + 1}} &= \bx_{t_l}^4 = \bx_{t_l}^3 + c_4\bv_{t_l}^3\Delta t, & \bv_{t_{l + 1}} &= \bv_{t_l}^4 = \bv_{t_l}^3,
\end{aligned}
\]
where the $(c_1, c_2, c_3, c_4)$ and $(d_1, d_2, d_3)$ are given as follows
\[
\begin{aligned}
w_0 = -\frac{\sqrt[3]{2}}{2 - \sqrt[3]{2}},  w_1 = \frac{1}{2 - \sqrt[3]{2}}, 
c_1 = c_4 = \frac{w_1}{2},  c_2 = c_2 = \frac{w_0 + w_1}{2}, 
d_1 = d_3 = w_1,  d_2 = w_0.
\end{aligned}
\]
\subsection{Estimating Planet Information}
Given only the position data of the celestial bodies, i.e., $\{\bx_i(t_l)\}_{i, l = 1}^{N, L}$ (indexed by table \ref{tab:CB_index}) for $T_0 = t_1 < \cdots < T_L = T$, observed daily ($t_2 = t_1 = 1$ day), we use the following algorithm to estimate the aphelion, perihelion, period and perihelion precession rates of the planets and the Moon. This algorithm is a simplified version of the idea presented in \cite{PFKWSZ2017} for calculating the perihelion precession rate of Mercury's orbit over a total of $2000$ years of trajectory data.  The basic idea behind our algorithm is that we treat the relative position between each planet and Sun (for the case of Moon, we take the relative position between Moon and Earth) as a continuous trajectory over time in space (hence the interpolation using splines over the whole observed time period).  We sample the time-points on such a trajectory at much refined intervals (i.e. at each minute) in order to obtain the relative distance data.  Then we use the physical definition of each term: aphelion/perihelion is at points which give the maximum/minimum relative distance, and the period is between each aphelion or perihelion points.  And for the calculation of the perihelion precession rates, we find the first time a planet/Moon reaches its perihelion, and then use it as the reference point, and record the positions of future perihelion points; then we compute the angle (also known as the precession advance) between each future perihelion point to the first perihelion point to obtain a sequence of precession advances (in time).  We fit a second-order polynomial time-series model to the series of precession advances, and use the coefficient for the first-order term as the precession rate.  Algorithm \ref{alg:ePI} shows the details.
\begin{algorithm}[H]
\caption{Estimating Planet Information}\label{alg:ePI}
\begin{algorithmic}[1]
\State Input: $\{\bx_i(t_l)\}_{i, l = 1}^{N, L}$.
\State Output: estimated aphelion, perihelion, period, and precession rate.
\For{\texttt{$i = 2, \ldots, N$}}
  \State{Calculate $\br_{1i}(t_l) = \bx_i(t_l) - \bx_1(t_l)$.}
  \State{Interpolate $br_{1i}(t_l)$ using splines, and obtain $\br_{1i}^{\text{spline}}$.}
  \State{Evaluate $\br_{1i}^{\text{spline}}$ at $t_l$ for $l = 1, \ldots, L'$ with $L' = 24 * 60 * L$.}
  \State{Calculate $r_{1i}^{\text{spline}}(t_l) = \norm{\br_{1i}^{\text{spline}}(t_l)}$.}
  \State{Find local maximum/minimum of the set $\{r_{1i}^{\text{spline}}(t_l)\}_{l = 1}^{L'}$.}
  \State{From $t_a$'s, where the local maximum of $\{r_{1i}^{\text{spline}}(t_l)\}_{l = 1}^{L'}$ are, find the mean/std of the aphelion.}
  \State{From $t_p$'s, where the local minimum of $\{r_{1i}^{\text{spline}}(t_l)\}_{l = 1}^{L'}$ are, find the mean/std of the perihelion.}
  \State{From $t_a$'s and $t_p$'s, find mean/std of the period.}
  \State{From $t_p$'s, find out the corresponding position vectors, i.e., $\br_{1i}^{\text{spline}}(t_p)$, and calculate the precession advances, $\theta_p$'s, from $p = 1$}.
  \State{Use the precession model fit, $\theta(t) = \beta_1 + \beta_2t + \beta_3t^2$ (according to \cite{PFKWSZ2017}), then $\beta_2$ will give an estimate of the precession rate.}
\EndFor
\end{algorithmic}
\end{algorithm}
\subsection{Estimating Masses of Celestial Bodies}
Recall the equations of motion which we assume for fitting our learning model to the observation data,
\[
\dot\bv_i(t) = \sum_{i' = 1}^N \intkernel_{i, i'}(\norm{\bx_{i'}(t) - \bx_i(t)})(\bx_{i'}(t) - \bx_i(t)), \quad i = 1, \cdots, N.
\]
Here the interaction kernel $\intkernel_{i, i'}: \R^+ \rightarrow \R$ is assumed to be either $\intkernel_{i, i} \equiv 0$ (when $i = i'$) or $\intkernel_{i, i'}(r) = \frac{U_{i, i'}(r)}{m_ir}$ (when $i \neq i'$) for some unknown gravitational potential $U_{i, i'}$, which might depend on $m_i$ and $m_{i'}$, possibly in a non-linear manner.  Although our learning method does not require any knowledge of the masses of the celestial bodies, the learned interaction kernels obtained from the observation of simple position, velocity data contain hidden information about the $m_i$s.  In order to gain insights into this additional structure of the estimators, we assume that 
\[
\lintkernel_{i, i'}(r) \approx \beta_{i'}f_m(r), \quad \text{for $i \neq i'$, and $i, i' = 1, \cdots, N$.}
\]
Then we use the non-linear de-coupling optimization based procedure first introduced in \cite{ZMM2020} in order to discover $\beta_i$s and $f_m$.  We improve the procedure so that it can tackle the ill-conditioning of the problem: the uneven mass distribution (the Sun is over $99\%$ of the mass of the Solar system), and that the CBs are at vast distances from each other.  

\begin{figure}[H]
\begin{subfigure}{\textwidth}
  \centering
   \includegraphics[width=0.7\textwidth]{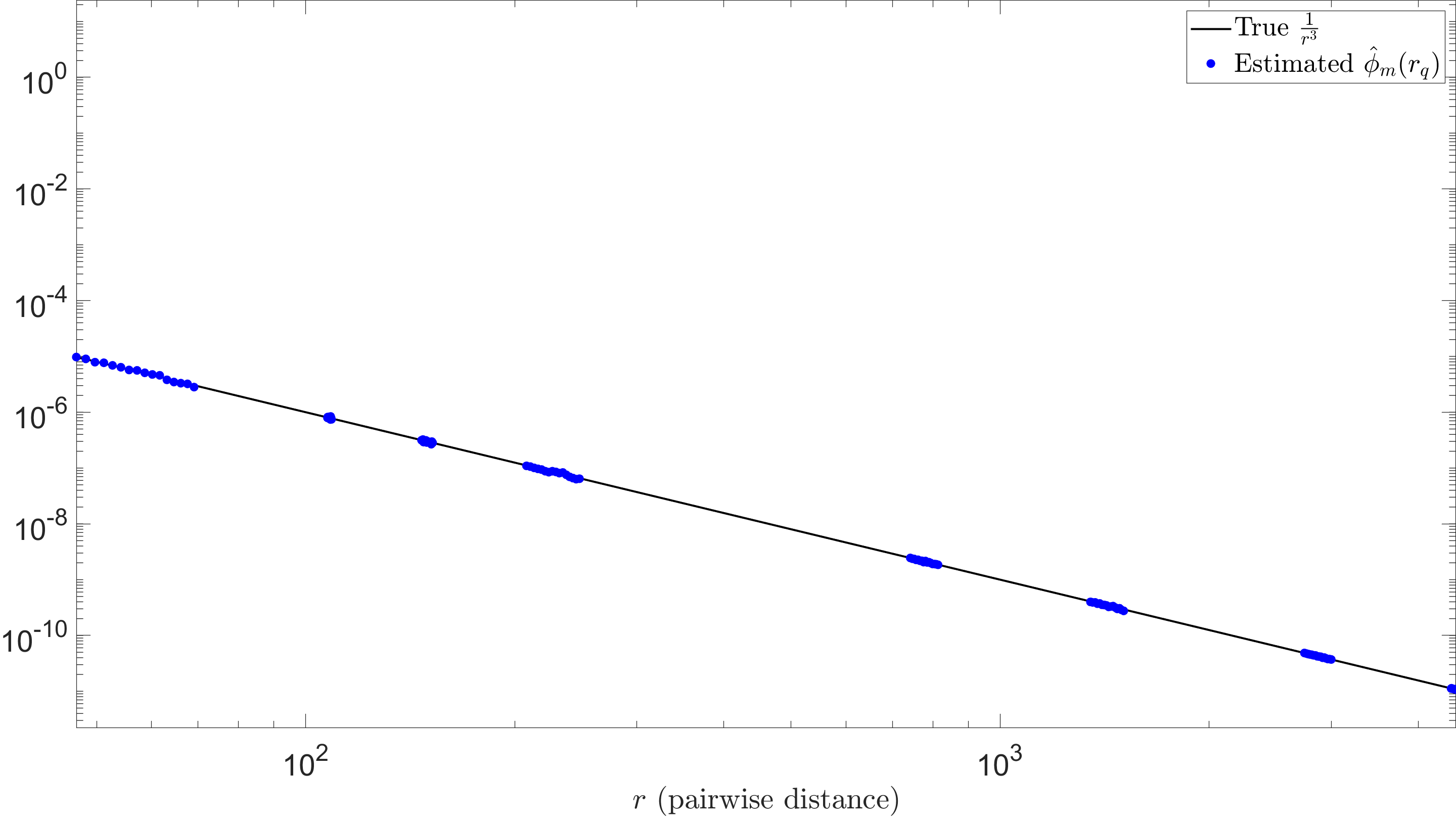} 
\end{subfigure}
\caption{Comparison of the discrete $\{f_m^*(r_q)\}_{q = 1}^Q$ to a continuous function $\frac{C}{r^3}$ for some $C > 0$.  We want to point out the vast distribution of pairwise distance for the CBs: the distance from Earth/Moon to Sun as overlapped with each other; starting from Mars, the gap between each CB has increased significantly, and yet, we are will able to de-coupled them as if they live on the function $\frac{C}{r^3}$.}
\label{fig:GSS_phim}
\end{figure}

\begin{figure}[H]
\centering
\includegraphics[width=0.7\textwidth]{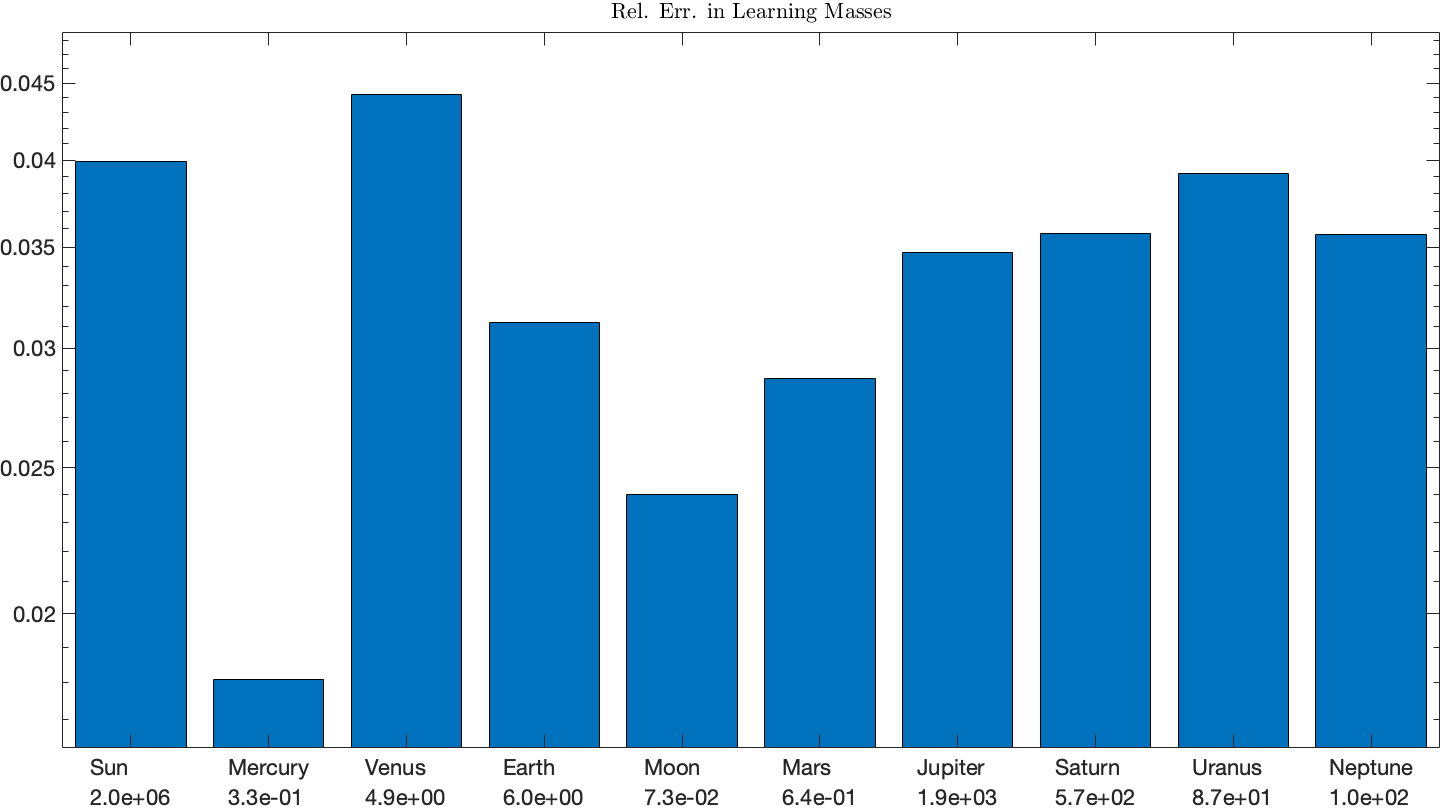} 
\caption{Relative Errors in Learning Masses.  True masses (taken from the JPL database) are shown in black text underneath the names of the corresponding CBs.   Relative errors are shown in blue error bars.  As shown, we have obtained highly accurate results from our de-coupling procedure with two-digit relative accuracy for most of the CBs, notwithstanding the ill-conditioning of the estimation procedure due to the high unevenness of the masses (the Sun takes over $99\%$ of the total mass of the system, and the CBs in the inner Solar system are rather small) and wide-spread distribution of CBs relative distance to the Sun.}
\label{fig:JPL_massErr_SI}
\end{figure}

Since $\lintkernel_{\idxcl, \idxcl'} \approx \intkernel_{\idxcl, \idxcl'}$ (for $\idxcl = 1$ or $\idxcl' = 1$), we will decouple $\beta_{\idxcl'}$ and $f_m(r)$ from $\lintkernel_{\idxcl, \idxcl'}$ through a two-step optimization procedure.  First, we consider a sequence of points $\{r_q\}_{q = 1}^Q$ from the supports of $\lintkernel_{1, \idxcl'}$ for $\idxcl' = 2, \cdots N$ ($r_q$'s are taken as the centers of the sub-intervals where the basis functions are built), and the following loss function,
\begin{align*}
\text{Loss}_1(\beta_1, f_m(r_1), \cdots, f_m(r_Q)) &= \sum_{\idxcl = 2}^N\sum_{q = 1}^Q\frac{(\lintkernel_{\idxcl, 1}(r_q) - \beta_1f_m(r_q))^2d\rho_T^{L, \idxcl, 1}(r_q)}{\sum_{q' = 1}^Q\lintkernel_{\idxcl, 1}(r_{q'})^2d\rho_T^{L, \idxcl, 1}(r_{q'})}
\end{align*}
Recall the distribution of pairwise distance data, $\rho_T^{L, i, i'}$, is given \eqref{eq:rho}.  Here $r_{i, i'}(t_l) = \norm{\bx_i(t_l) - \bx_{i'}(t_l)}$ and $\rho_T^{L, i, i'}$ is computed over the observation data, $\{\bx_i(t_l), \bv_i(t_l)\}_{i, l = 1}^{N, L}$.  $\text{Loss}_1$ is minimized over $\beta_{1} \ge 0$ and $f_m(r_q) \in R$ for $q = 1, \cdots, Q$.  The purpose of using a scaled version of the difference is to reduce the effect of distance on the CBs' interaction kernels.  The result for learning the discrete $\{f_m^*(r_q)\}_{q = 1}^Q$ is shown in figure \ref{fig:GSS_phim}.
Next, we use $\{f_m(r_q)\}_{q = 1}^Q$ to learn the $\beta_{\idxcl}$ for $\idxcl = 2, \ldots, N$, via the following loss function,
\begin{align*}
\text{Loss}_2(\beta_2, \cdots, \beta_N) &= \sum_{\idxcl' = 2}^N\sum_{q = 1}^Q\frac{(\lintkernel_{1, \idxcl'}(r_q) - \beta_{\idxcl'}f_m^*(r_q))^2d\rho_T^{L, 1, \idxcl'}(r_q)}{\sum_{q' = 1}^Q(\lintkernel_{1, \idxcl'}(r_{q'}))^2d\rho_T^{L, 1, \idxcl'}(r_{q'})}
\end{align*}
The scaling is to reduce the effect of uneven distribution of mass of each CB as well as the vast distribution of relative distance.  The appropriate use of the dynamics-adapted measures enables us to identify parameters correctly. $\text{Loss}_2$ is minimized over $\beta_{\idxcl'} \ge 0$ for $\idxcl' = 2, \cdots, N$.  Furthermore by assuming that $f_m(r) = \frac{C}{r^p}$, we found that $p = 3$ gives the best fit, and also discover the masses of the celestial bodies.  Results are shown in the main body of this work. These steps are taken only in order to estimate the masses; the originally estimated interaction kernels $\lintkernel_{i, i'}$ are those used to produce estimated and predicted trajectories, and all the other results (periods, aphelia, precession rates, etc...).  In order to gain deeper sight of the difficulty of estimating the masses of each CB, we provide the relative errors for estimating the mass of each astronomical object in figure \ref{fig:JPL_massErr_SI} along with the ``true'' masses provided by NASA.
\section{Learning Results}
\begin{figure}[H]
\centering
\begin{subfigure}[b]{0.31\textwidth}
\centering
\includegraphics[width=\textwidth]{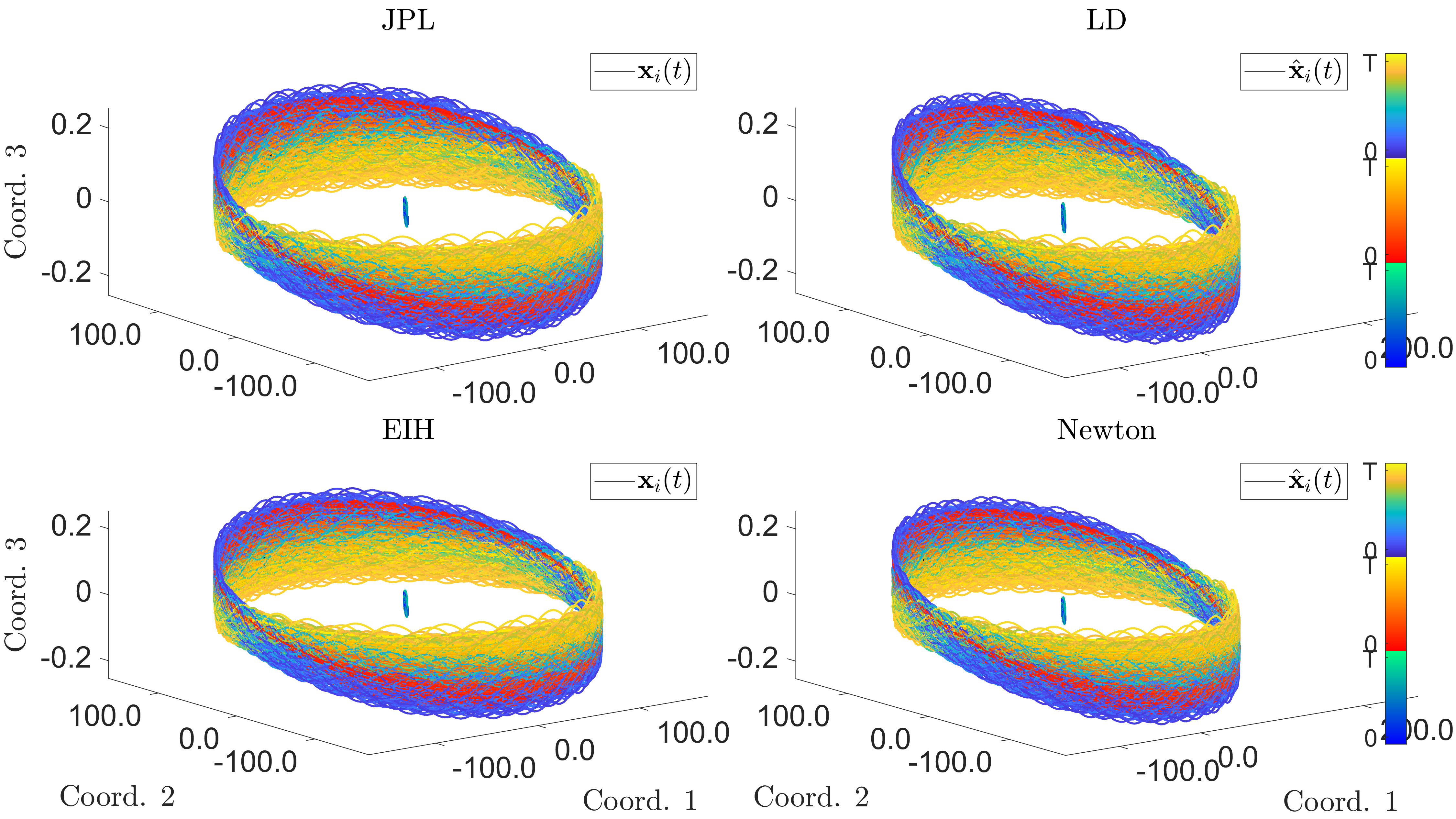}
\caption{Earth-Moon-Sun}
\end{subfigure} ~
\begin{subfigure}[b]{0.31\textwidth}
\centering
\includegraphics[width=\textwidth]{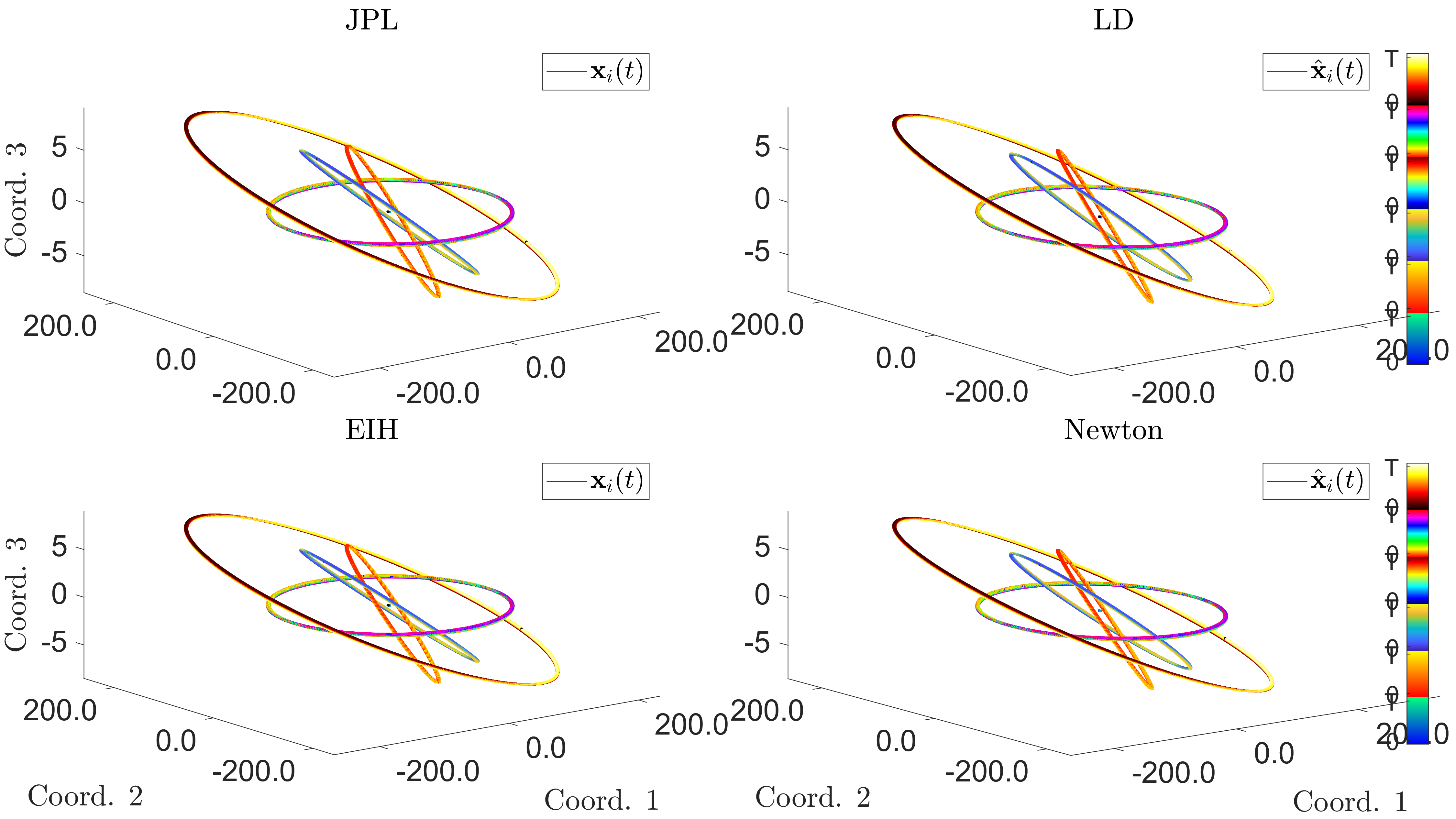}
\caption{Inner Solar}
\end{subfigure} ~
\begin{subfigure}[b]{0.31\textwidth}
\centering
\includegraphics[width=\textwidth]{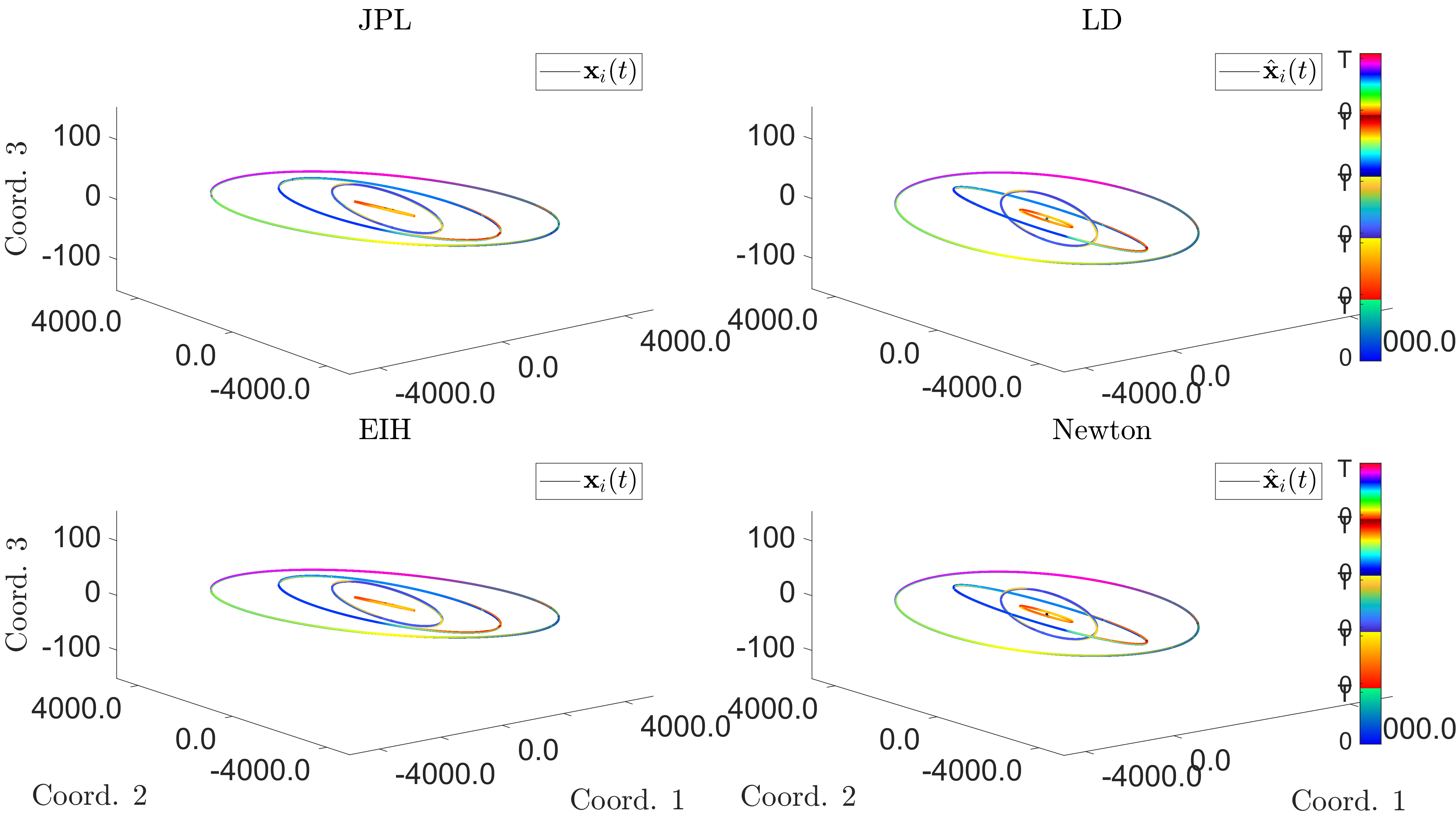}
\caption{Outer Solar}
\end{subfigure}
\caption{Comparison of $4$ different models: JPL, LD, EIH, and Newton for $3$ different types of sub-solar system: Earth-Moon-Sun, inner solar system and outer solar system.}
\label{fig:JPL_trajComp}
\end{figure}
In this section, we present the detailed setup for the learning experiments conducted on $500$ years of daily sampled position/velocity data (from year $1500$ to $1999$) of $10$ CBs, and their indices are given in table \ref{tab:CB_index}.
\begin{table}[H]
\centering
 \begin{tabular}{ c | c | c | c | c | c } 
 \hline 
 CB    & Sun & Mercury & Venus & Earth & Moon  \\
 \hline
 Index & $1$ & $2$     & $3$   & $4$   & $5$ \\
 \hline
 \hline
 CB    & Mars & Jupiter & Saturn & Uranus & Neptune \\
 \hline
 Index & $6$  &  $7$    & $8$    & $9$    &$10$    \\
 \hline
 \end{tabular}
 \caption{Indexing of $10$ CBs.}
 \label{tab:CB_index}
\end{table}
We conform to the units used in NASA's measures, hence we use $10^{24}$ kg for the unit of mass, $10^{6}$ km for the unit of length, and $1$ day for the unit of time, the values for the gravitational constant ($G$) and speed of light ($c$) have to be re-scaled, see table \ref{tab:constants}.
\begin{table}[H]
\centering
 \begin{tabular}{ c | c} 
 \hline
 $G$  & $c$  \\
 \hline
 $4.98217402368 \cdot 10^{-4} \, \frac{(10^6 \text{km})^3}{10^{24} \text{kg} \cdot (\text{day})^3}$ & $2.59020683712 \cdot 10^{4} \, \frac{10^6 \text{km}}{\text{day}}$\\
 \hline
 \end{tabular}
 \caption{Important Constants}
 \label{tab:constants}
\end{table}
In order to compute the relative errors of the estimation of masses of CBs from our learning algorithm, we use the values for the mass of each CB in table \ref{tab:JPL_params}.
\begin{table}[H]
\centering
 \begin{tabular}{ c | c | c | c | c | c } 
 \hline 
 CB   & Sun & Mercury & Venus & Earth & Moon  \\
 \hline
 Mass &  $1.9885 \cdot 10^6$   &   $0.330$             &      $4.87$       &      $5.97$       &      $0.073$         \\
 \hline
 \hline
 CB   & Mars & Jupiter & Saturn & Uranus & Neptune \\
 \hline
 Mass & $0.642$       &  $1898$            &      $568$        &        $86.8$        &$102$    \\
 \hline
 \end{tabular}
 \caption{Masses of CBs, unit $10^{24}$ kg.}
 \label{tab:JPL_params}
\end{table}
We perform various learning experiments on a computing workstation provided by Prisma Analytics, Inc.  It has $2$ Intel Xeon $E5$-$2687W$ CPUs, each with $12$ computing cores, $512$ GB memory, and runs on Ubuntu $16.04.7$ LTS operating system.   The parallel environment is implemented in MATLAB with the ``parfor'' command.  We found that, having implemented the parallelization routine, the computational bottleneck comes from the long-time symplectic integration, which was expected.

\textbf{Choosing $(SI_{i, i'}, p_{i, i'})_{i,i'}$}: we conduct various experiments at finding the best $(S_{i, i'}, p_{i, i'})$ combination (here $S_{i, i'}$ stands for the number of sub-intervals for $[R_{i, i'}^{\min},  R_{i, i'}^{\max}]$ and $p_{i, i'}$ is the degree of the Clamped B-spline basis for estimating $\intkernel_{i, i'}$) in terms of trajectory error estimation as well as perihelion precession rate estimation for the Mercury's orbit (we taken the Effective Theory's approach).  The combination of $(S_{i, i'}, p_{i, i'}) = (90, 4)$ gives the best performance.  

\textbf{EIH Range}: In order to make the comparison more meaningful,  we compare our learned $\lintkernel_{i, i'}$ to Newton's gravity and the EIH forces for each $(i, i')$ pair.  We define 
\[
\text{Newton}_{i, i'}(r) = \frac{Gm_{i'}}{r^3};
\]
for the EIH forces, we need to find out the range of the projection, we first obtain the $\dot\bv_i(t)$ given $\{\bx_i(t), \bv_i(t)\}_{i = 1}^N$ via \eqref{eq:EIH_eq}, then obtain (over time) the EIH force projected on to $\bx_{i'}(t) - \bx_i(t)$ as follows
\[
\begin{aligned}
\text{EIH}_{i, i'}(t) &= \frac{Gm_{i'}}{\norm{\bx_{i'} - \bx_{i}}^3}\cdot(\bx_{i'} - \bx_i)\Bigg\{1 - \frac{4}{c^2}\sum_{\substack{i'' = 1 \\ i'' \neq i}}^{N}\frac{Gm_{i''}}{\norm{\bx_{i''} - \bx_i}} -\frac{1}{c^2}\sum_{\substack{i'' = 1 \\ i'' \neq i'}}^{N}\frac{Gm_{i''}}{\norm{\bx_{i''} - \bx_{i'}}} \nonumber\\
  &\quad + \frac{2\norm{\bv_{i'} - \bv_i}^2 - \norm{\bv_i}^2}{c^2} - \frac{3}{2c^2}\big(\dotp{\frac{\bx_{i'} - \bx_{i}}{\norm{\bx_i - \bx_{i'}}}, \bv_{i'}}\big)^2 + \frac{1}{2c^2}\dotp{\bx_{i'} - \bx_{i}, \dot\bv_{i'}}\Bigg\}
\end{aligned},
\]
notice that the dependence of $\bx_i$ and $\bv_i$ on $t$ is suppressed to simplify the notation.  We then compute statistics of $\text{EIH}_{i, i'}$ over $[R_{i, i'}^{\min},  R_{i, i'}^{\max}]$ to obtain the maximum/minimum, i.e. . $\text{EIH}_{i, i'}^{\min, \max}(r)$.  Then we consider the following relative errors,
\[
\text{Err}_{i, i'}^1(r) = \frac{\text{EIH}_{i, i'}^{\max}(r) - \text{Newton}_{i, i'}(r)}{\text{Newton}_{i, i'}(r)} \quad \text{and} \quad\text{Err}_{i, i'}^2(r) = \frac{\text{EIH}_{i, i'}^{\min}(r) - \text{Newton}_{i, i'}(r)}{\text{Newton}_{i, i'}(r)}
\]
and
\[
\text{Err}_{i, i'}^3(r) = \frac{\lintkernel_{i, i'}(r) - \text{Newton}_{i, i'}(r)}{\text{Newton}_{i, i'}(r)}.
\]

\begin{figure}[H]
\centering
\begin{subfigure}[b]{0.48\textwidth} 
\centering
\includegraphics[width=\textwidth]{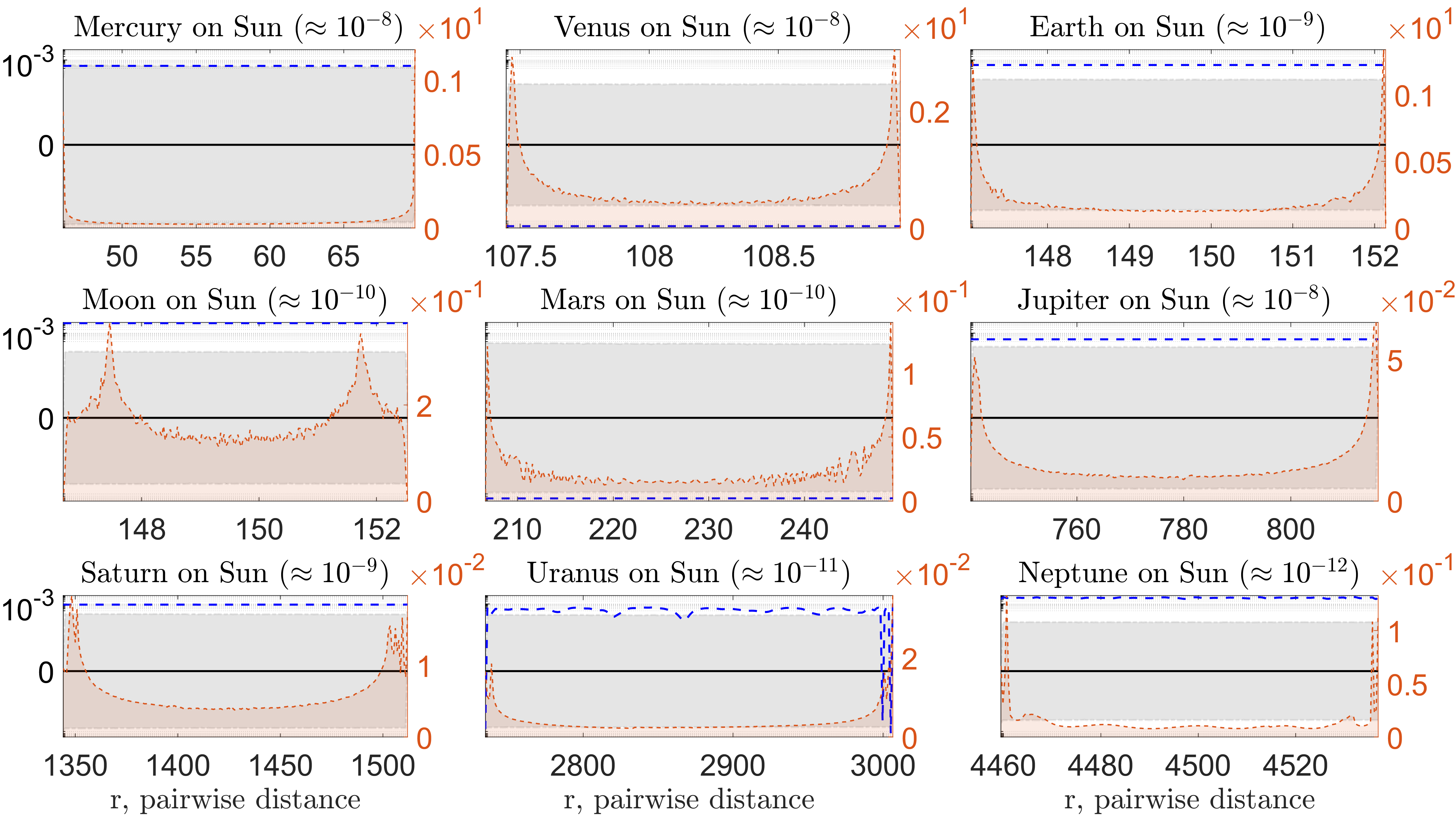} 
\end{subfigure} ~
\begin{subfigure}[b]{0.48\textwidth}
\centering
\includegraphics[width=\textwidth]{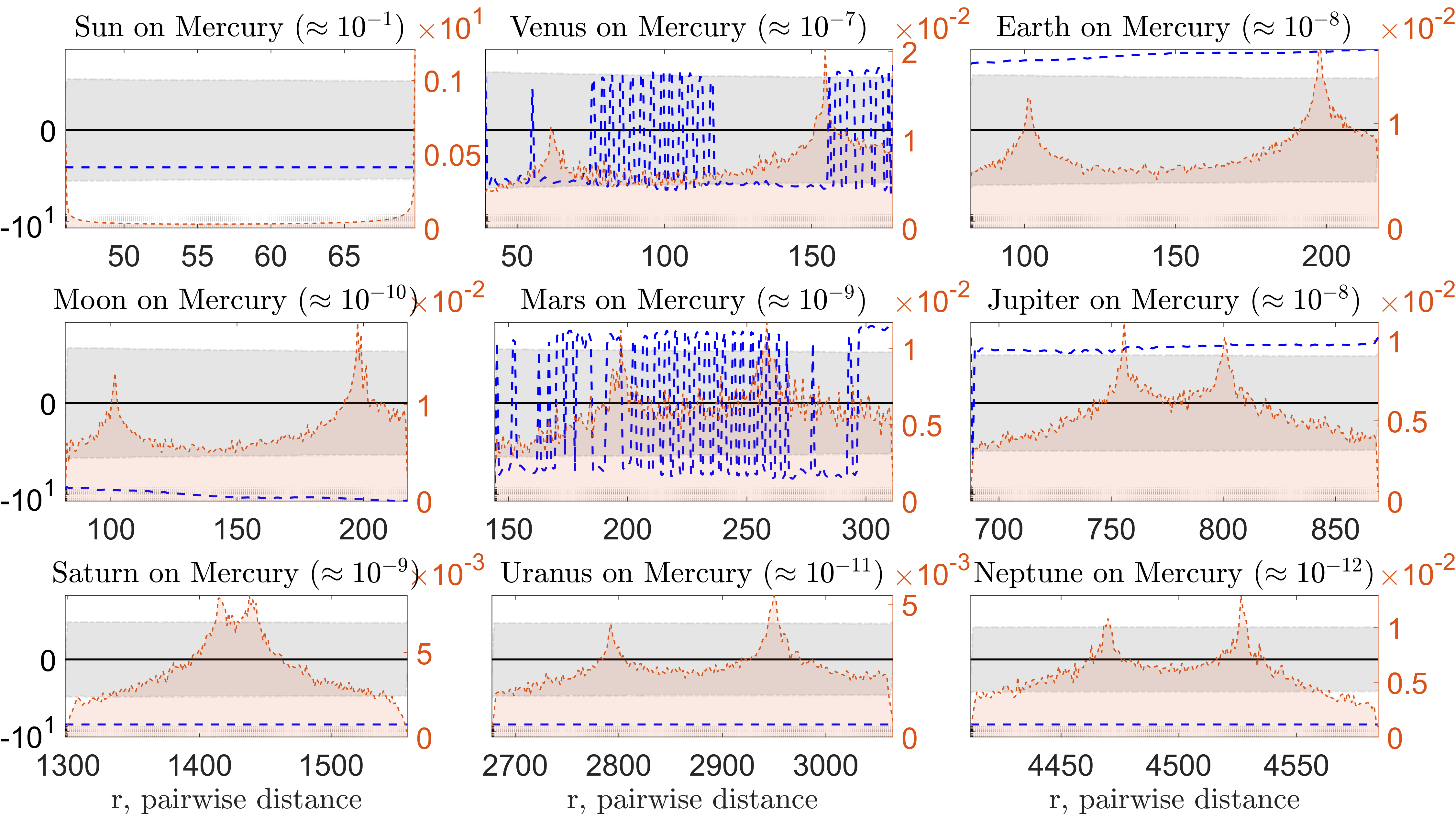}
\end{subfigure}
\caption{CB-on-Sun ($(\lintkernel_{1, i})_i$) and CB-on-Mercury ($(\lintkernel_{2, i})_i$) interaction kernels vs. Newton's gravity and the EIH range, shown in terms of relative error compared to Newton in symmetric-log scale.  Shown in the background are the corresponding distribution of pairwise distance data used to estimate these kernels, i.e. $\rho_{T, i, i'}^{L}$.  Dotted black lines represent the errors $\text{Err}_{i, i'}^1(r)$ and $\text{Err}_{i, i'}^2(r)$ for $i = 1, 2$.  Solid line black line represents the error for Newton's gravity, which is exactly zero.  Solid blue line shows the error $\text{Err}_{i, i'}^3(r)$ for $i = 1, 2$.  As shown in the each sub-plot, our learned interaction kernels are recovered in a way which is within the EIH range and close to Newton's gravity.}
\label{fig:JPL_phiEhats_12}
\end{figure}

\begin{figure}[H]
\centering
\begin{subfigure}[b]{0.48\textwidth} 
\centering
\includegraphics[width=\textwidth]{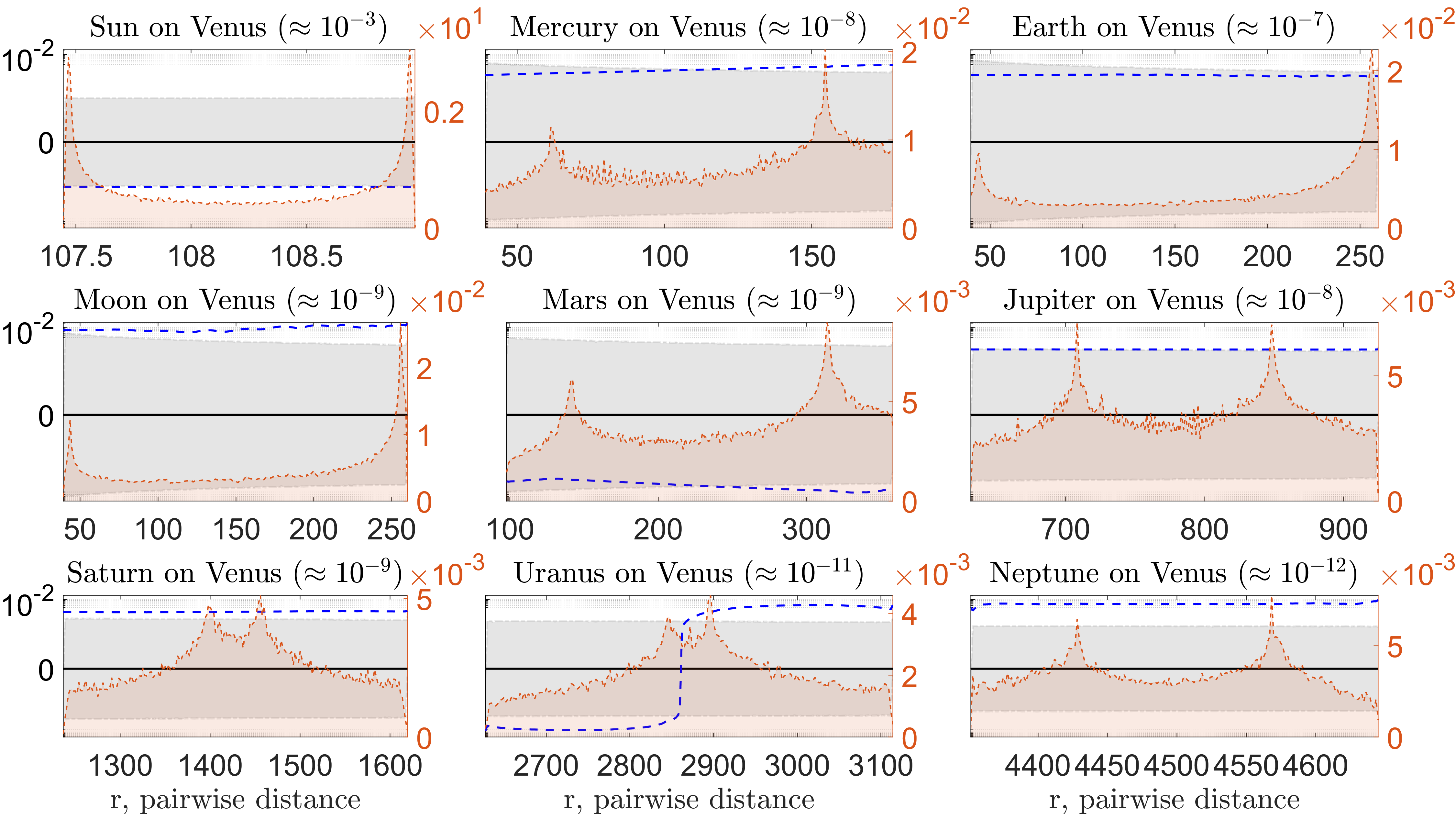} 
\end{subfigure} ~
\begin{subfigure}[b]{0.48\textwidth}
\centering
\includegraphics[width=\textwidth]{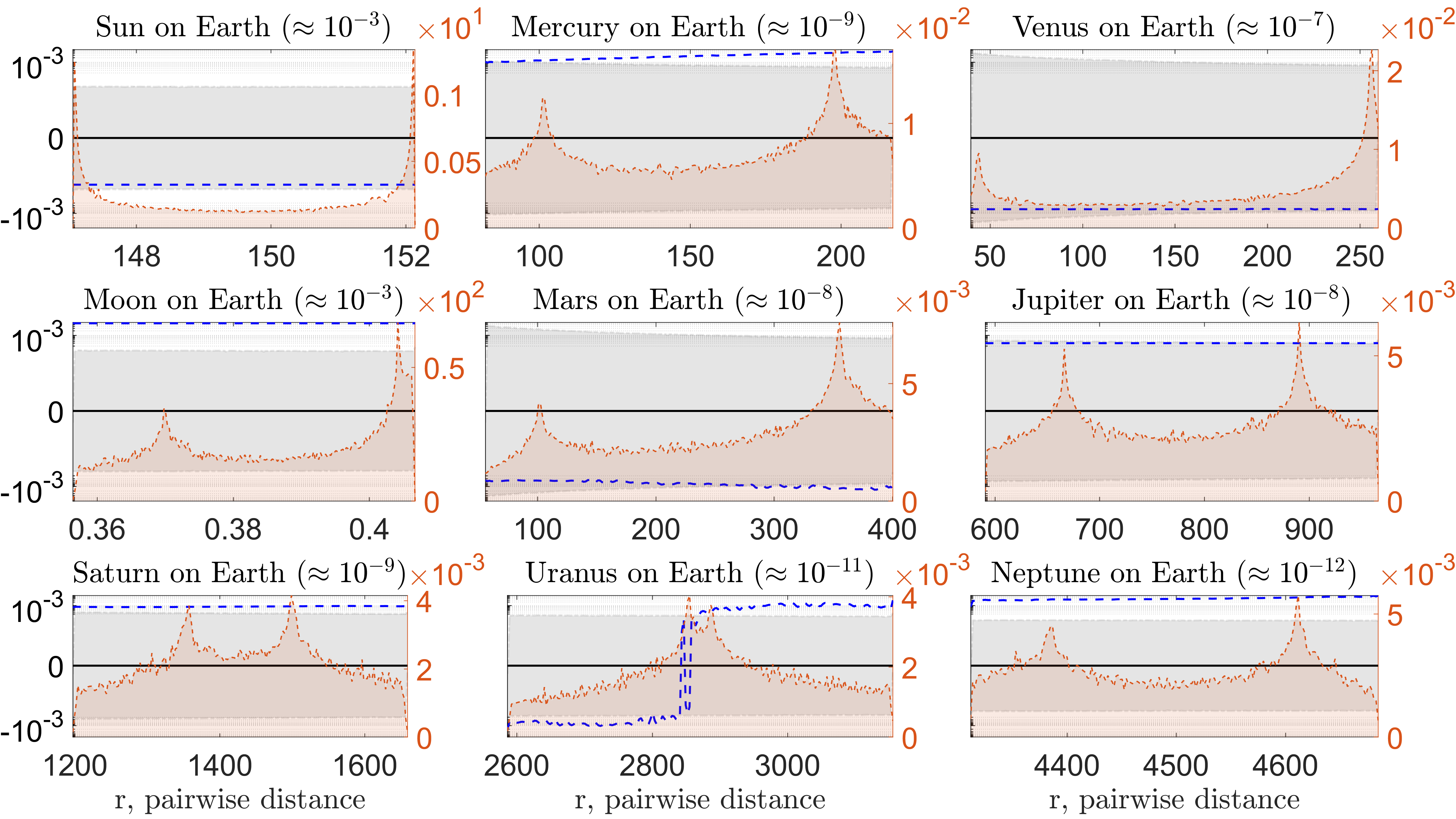}
\end{subfigure}
\caption{CB-on-Venus ($(\lintkernel_{3, i})_i$) and CB-on-Earth ($(\lintkernel_{4, i})_i$) interaction kernels vs. Newton's gravity and the EIH range, shown in terms of relative error compared to Newton in symmetric-log scale.  Shown in the background are the corresponding distribution of pairwise distance data used to estimate these kernels, i.e. $\rho_{T, i, i'}^{L}$.  Dotted black lines represent the errors $\text{Err}_{i, i'}^1(r)$ and $\text{Err}_{i, i'}^2(r)$ for $i = 3, 4$.  Solid line black line represents the error for Newton, which is exactly zero.  Solid blue line shows the error $\text{Err}_{i, i'}^3(r)$ for $i = 3, 4$.  As shown in the each sub-plot, our learned interaction kernels are recovered in a way which is within the EIH range and close to Newton's gravity.}
\label{fig:JPL_phiEhats_34}
\end{figure}

\begin{figure}[H]
\centering
\begin{subfigure}[b]{0.48\textwidth} 
\centering
\includegraphics[width=\textwidth]{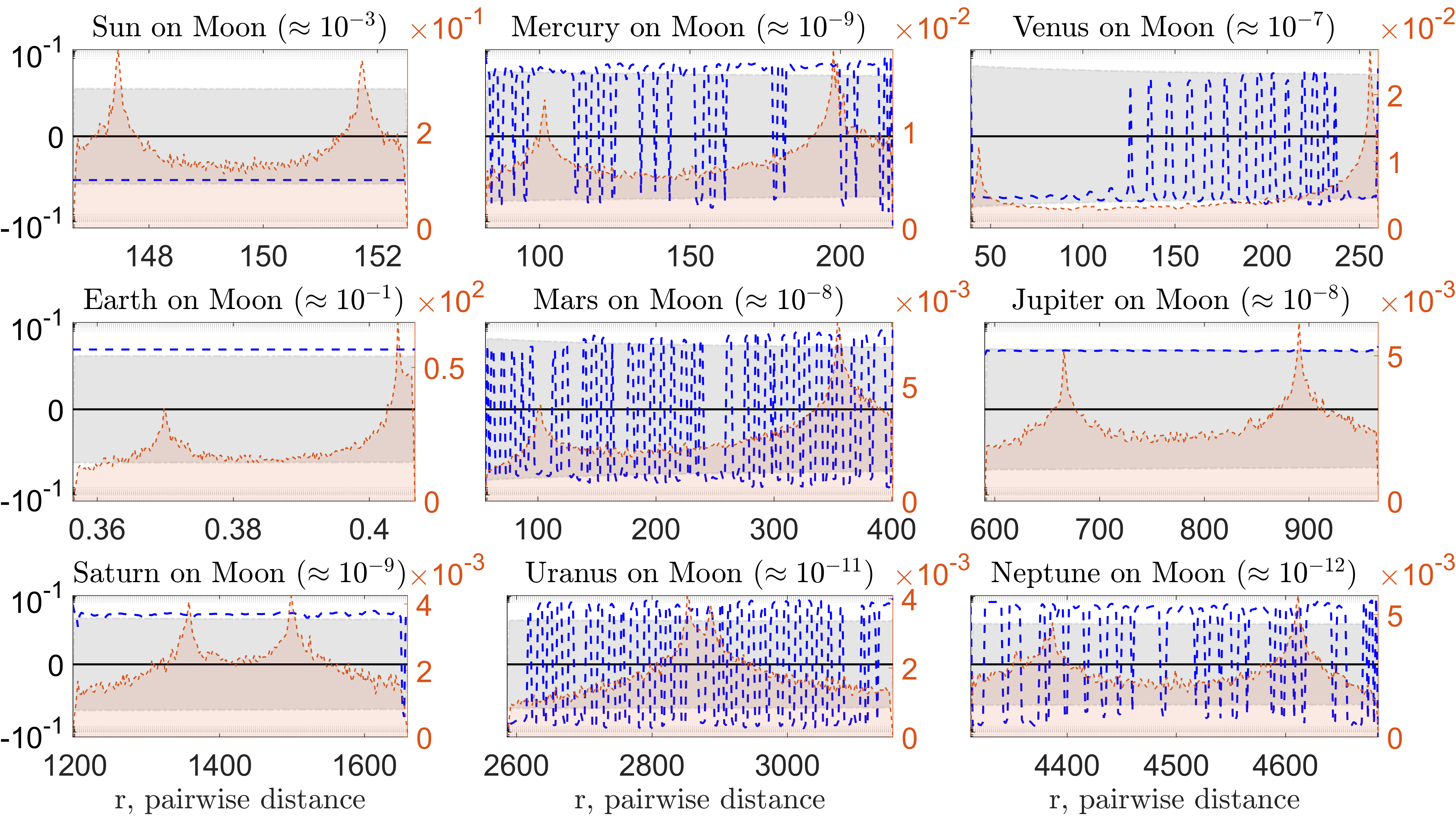} 
\end{subfigure} ~
\begin{subfigure}[b]{0.48\textwidth}
\centering
\includegraphics[width=\textwidth]{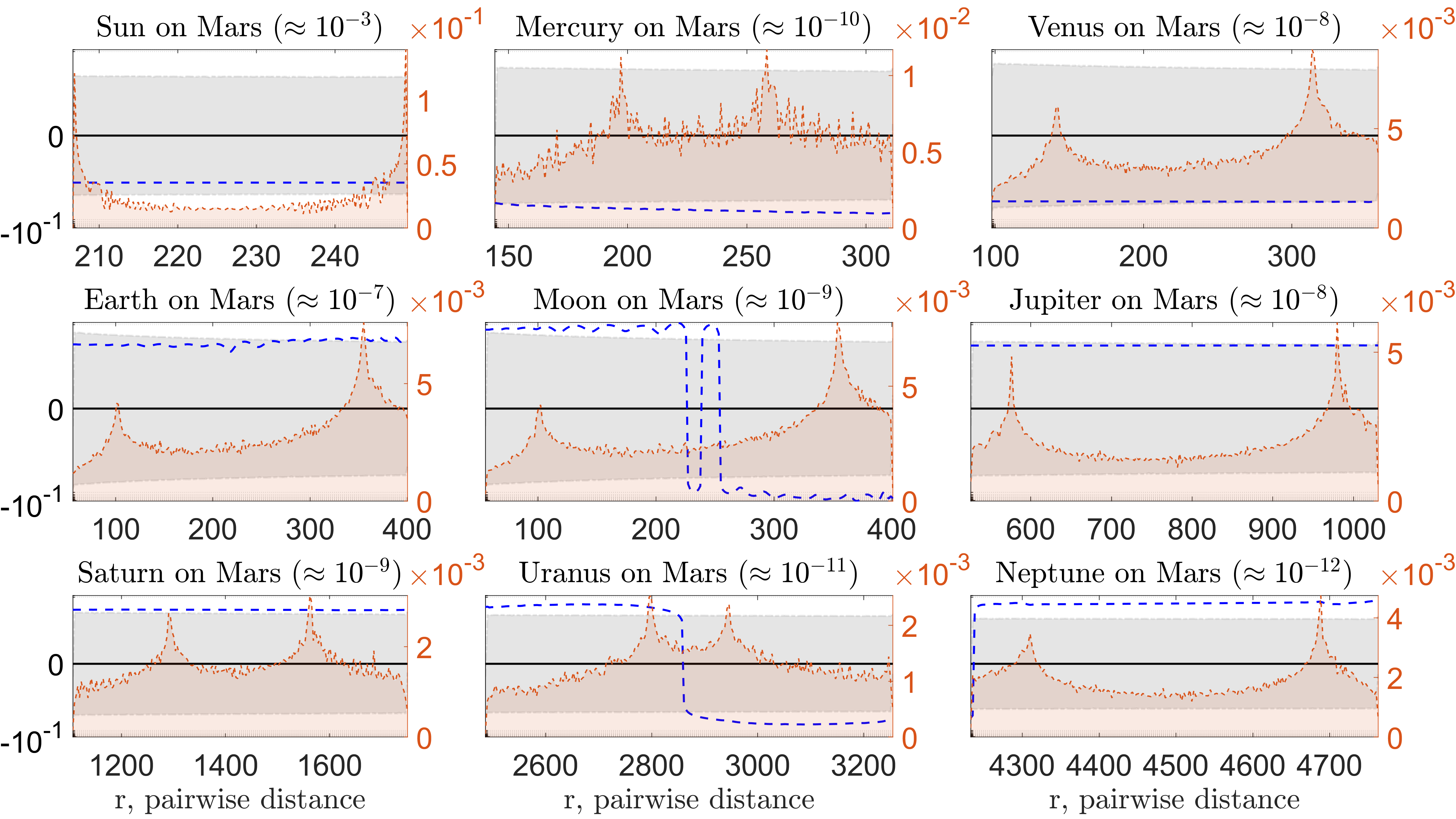}
\end{subfigure}
\caption{CB-on-Moon ($(\lintkernel_{5, i})_i$) and CB-on-Mars ($(\lintkernel_{6, i})_i$) interaction kernels vs. Newton's gravity and the EIH range, shown in terms of relative error compared to Newton in symmetric-log scale.  Shown in the background are the corresponding distribution of pairwise distance data used to estimate these kernels, i.e. $\rho_{T, i, i'}^{L}$.  Dotted black lines represent the errors $\text{Err}_{i, i'}^1(r)$ and $\text{Err}_{i, i'}^2(r)$ for $i = 5, 6$.  Solid line black line represents the error for Newton, which is exactly zero.  Solid blue line shows the error $\text{Err}_{i, i'}^3(r)$ for $i = 5, 6$.  As shown in the each sub-plot, our learned interaction kernels are recovered in a way which is within the EIH range and close to Newton's gravity.}
\label{fig:JPL_phiEhats_56}
\end{figure}

\begin{figure}[H]
\centering
\begin{subfigure}[b]{0.48\textwidth} 
\centering
\includegraphics[width=\textwidth]{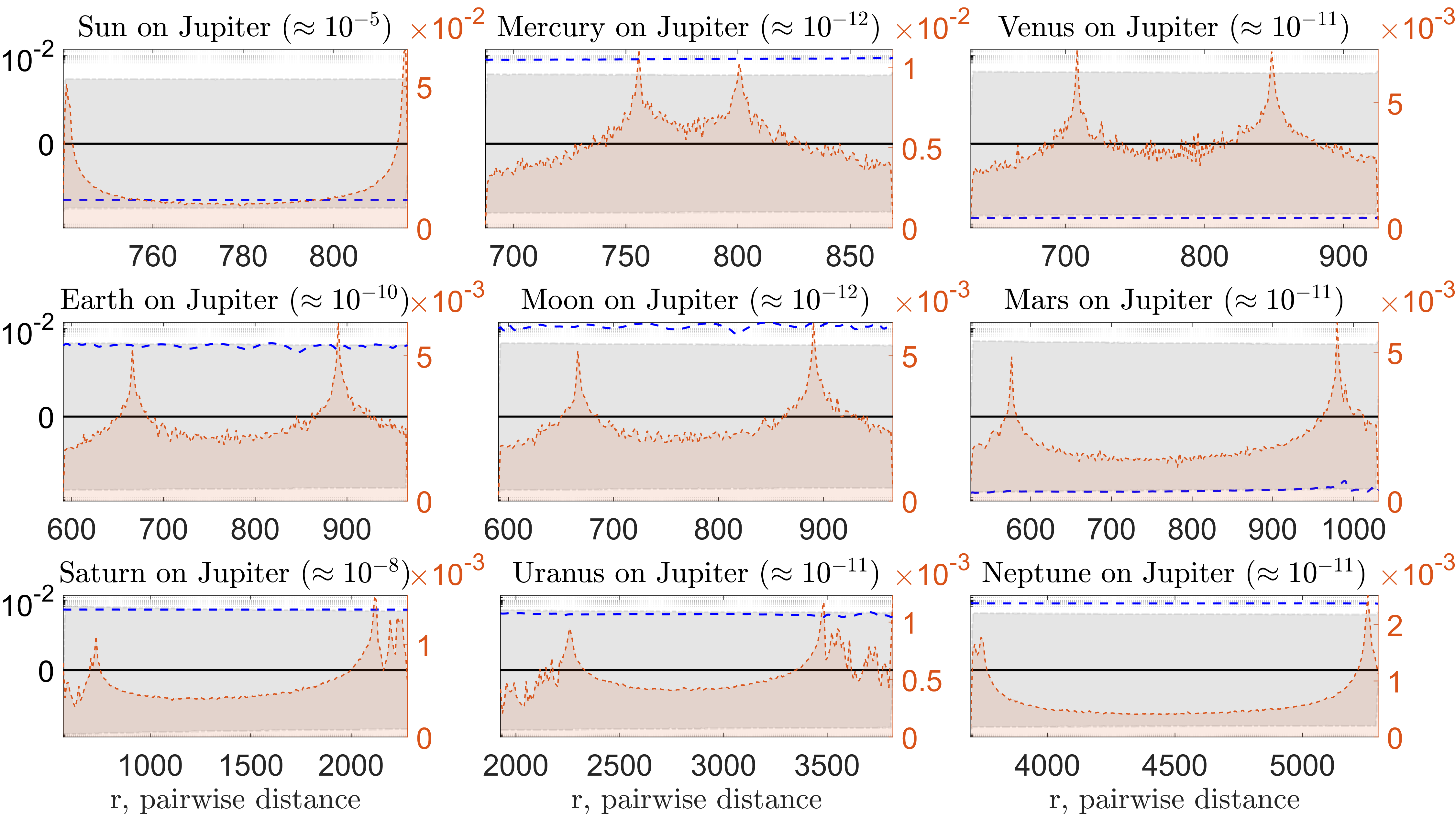} 
\end{subfigure} ~
\begin{subfigure}[b]{0.48\textwidth}
\centering
\includegraphics[width=\textwidth]{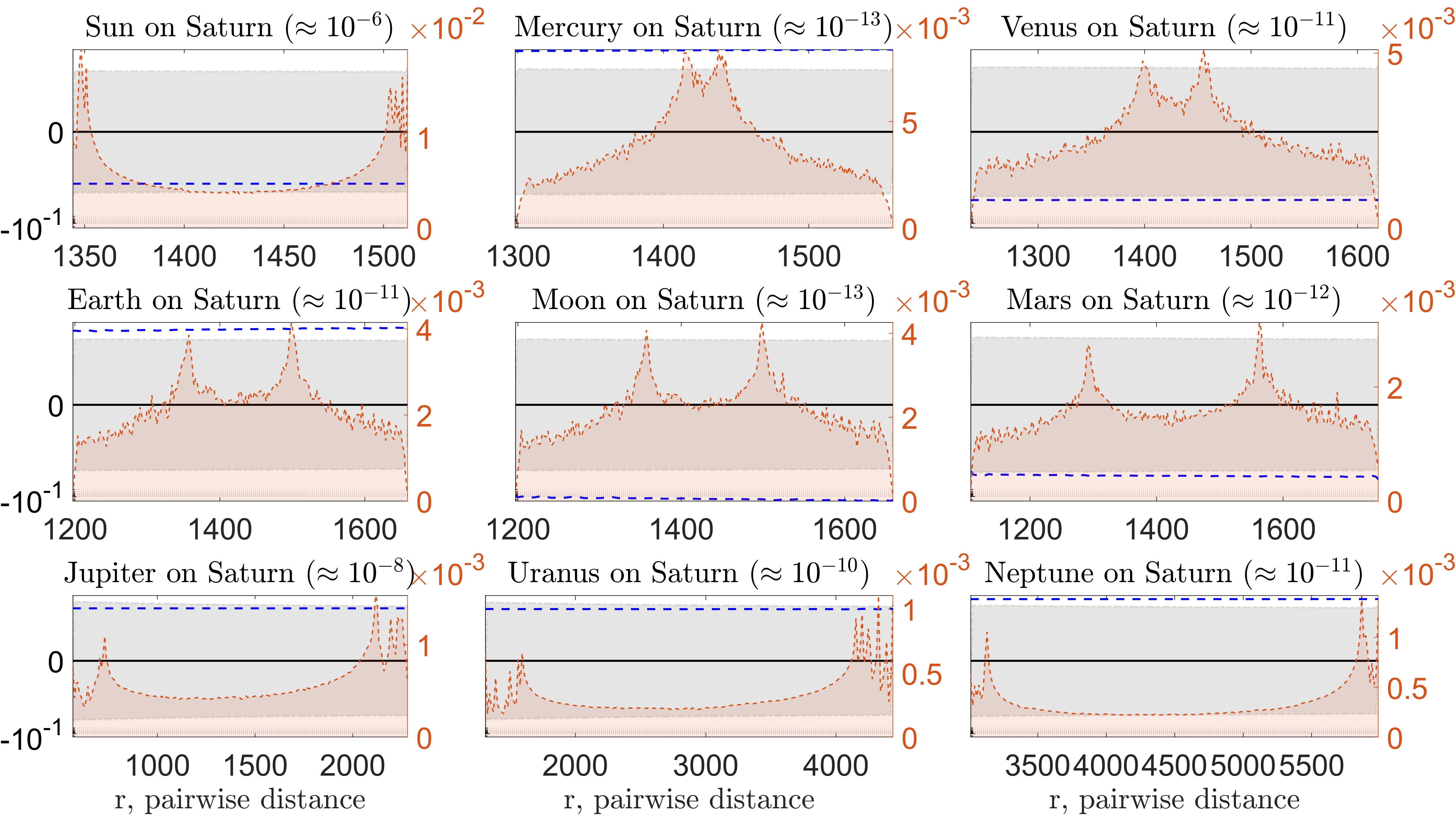}
\end{subfigure}
\caption{CB-on-Jupiter ($(\lintkernel_{7, i})_i$) and CB-on-Saturn ($(\lintkernel_{8, i})_i$) interaction kernels vs. Newton's gravity and the EIH range, shown in terms of relative error compared to Newton in symmetric-log scale.  Shown in the background are the corresponding distribution of pairwise distance data used to estimate these kernels, i.e. $\rho_{T, i, i'}^{L}$.  Dotted black lines represent the errors $\text{Err}_{i, i'}^1(r)$ and $\text{Err}_{i, i'}^2(r)$ for $i = 7, 8$.  Solid line black line represents the error for Newton, which is exactly zero.  Solid blue line shows the error $\text{Err}_{i, i'}^3(r)$ for $i = 7, 8$.  As shown in the each sub-plot, our learned interaction kernels are recovered in a way which is within the EIH range and close to Newton's gravity.}
\label{fig:JPL_phiEhats_78}
\end{figure}

\begin{figure}[H]
\centering
\begin{subfigure}[b]{0.48\textwidth}
\centering
\includegraphics[width=\textwidth]{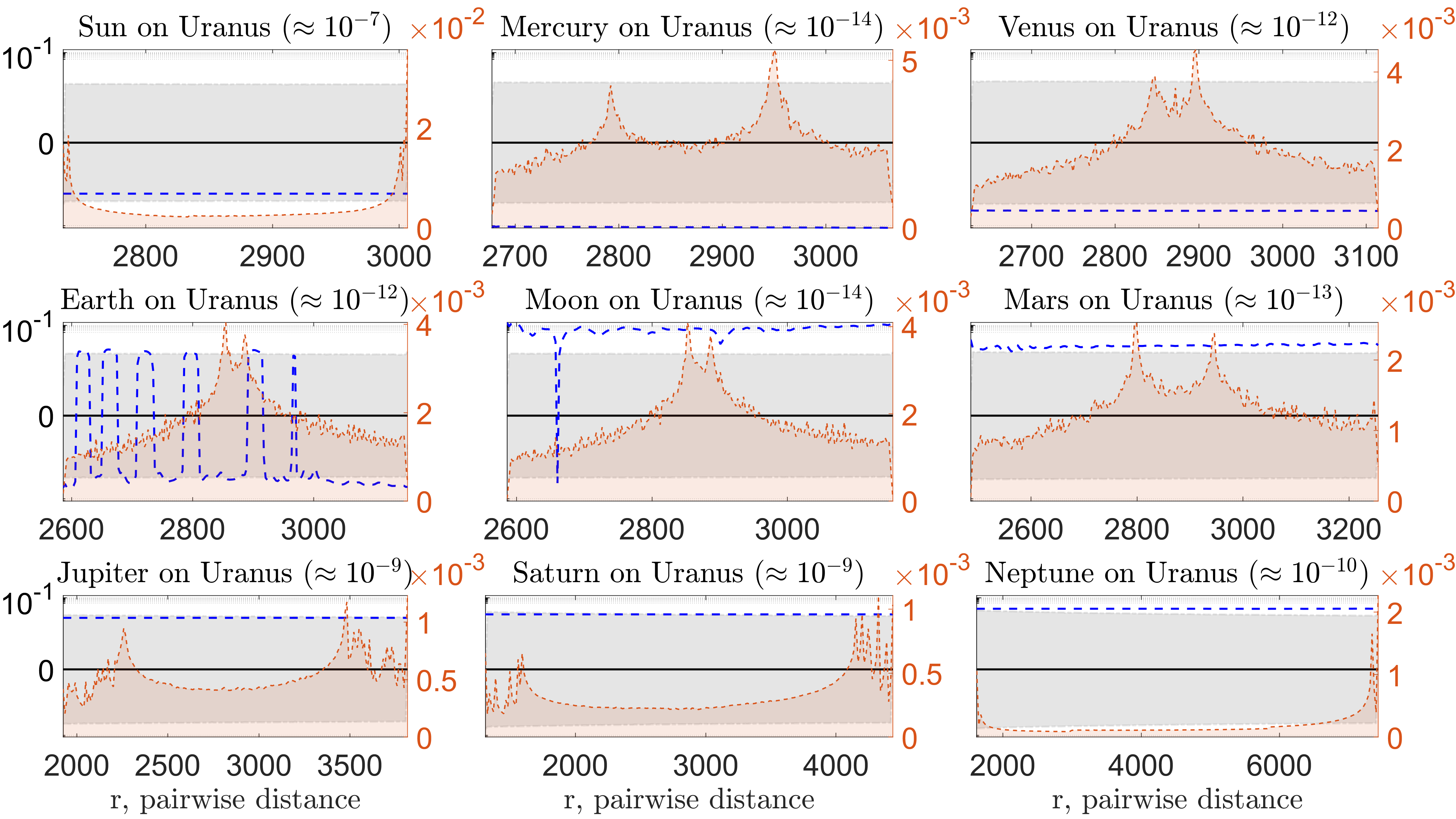}
\end{subfigure} ~
\begin{subfigure}[b]{0.48\textwidth}
\centering
\includegraphics[width=\textwidth]{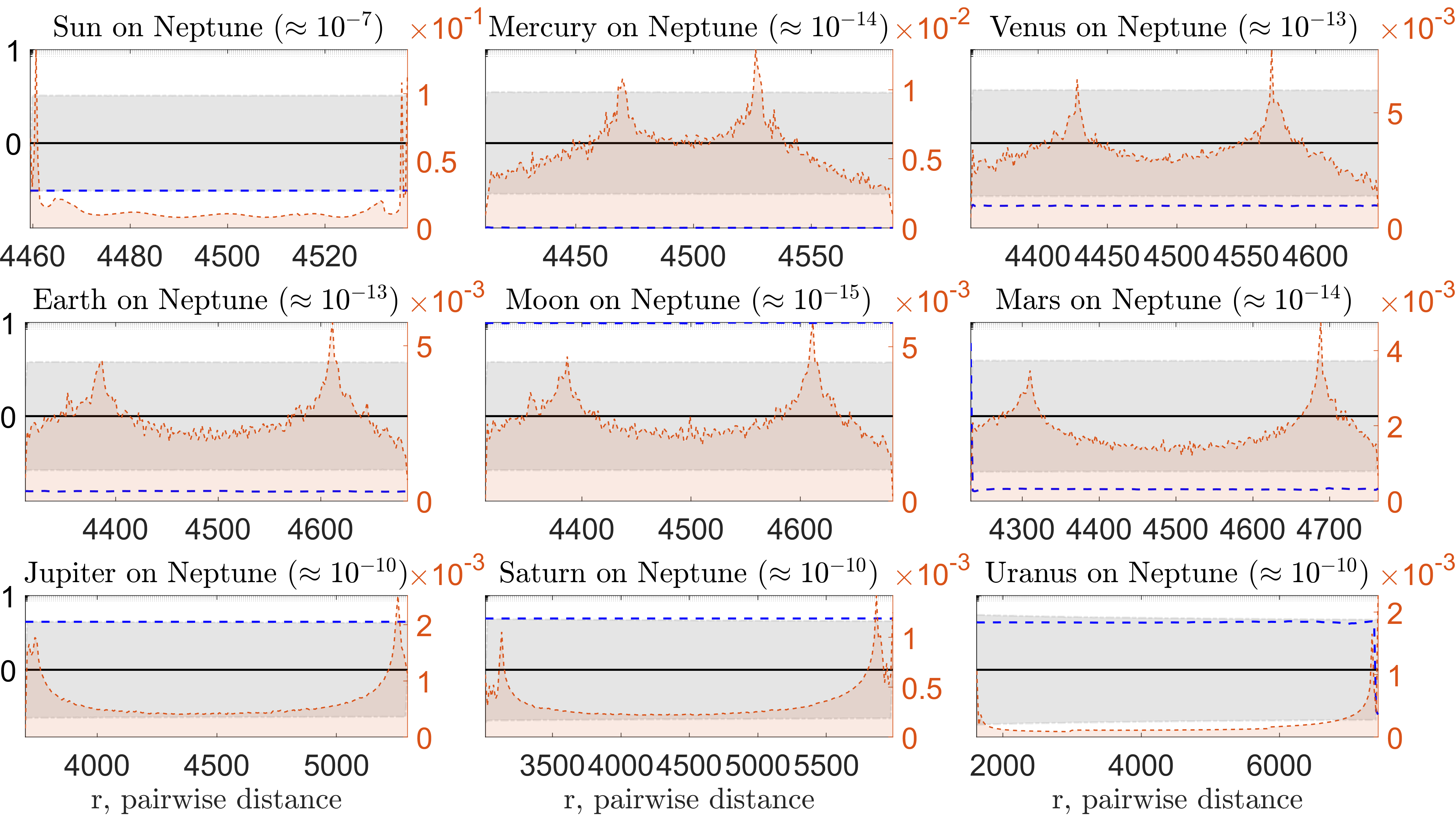}
\end{subfigure}
\caption{CB-on-Uranus ($(\lintkernel_{9, i})_i$) and CB-on-Neptune ($(\lintkernel_{10, i})_i$) interaction kernels vs. Newton's gravity and the EIH range, shown in terms of relative error compared to Newton in symmetric-log scale.  Shown in the background are the corresponding distribution of pairwise distance data used to estimate these kernels, i.e. $\rho_{T, i, i'}^{L}$.  Dotted black lines represent the errors $\text{Err}_{i, i'}^1(r)$ and $\text{Err}_{i, i'}^2(r)$ for $i = 9, 10$.  Solid line black line represents the error for Newton, which is exactly zero.  Solid blue line shows the error $\text{Err}_{i, i'}^3(r)$ for $i = 9, 10$.  As shown in the each sub-plot, our learned interaction kernels are recovered in a way which is within the EIH range and close to Newton's gravity.}
\label{fig:JPL_phiEhats_910}
\end{figure}

\textbf{Comparison of $\lintkernel_{i, i'}$'s}: we present the comparisons for $\lintkernel_{i, i'}$s for $i = 3, 4, 7, 8, 9, 10$ with the special relative errors defined above, and show them in symmetric log scale so that the details towards zero can be shown properly; it is done via a special transformation which guarantees the continuity across zero \cite{Webber2012}.  The results are shown in figures \ref{fig:JPL_phiEhats_12}, \ref{fig:JPL_phiEhats_34}, \ref{fig:JPL_phiEhats_56}, \ref{fig:JPL_phiEhats_78}, and \ref{fig:JPL_phiEhats_910}.

%
\nocite{*}

\bibliography{gravity_letter}

\end{document}